\documentclass[aps,amsmath,amssymb,prd,showpacs,floatfix,preprint,superscriptaddress,nofootinbib,12pt]{article}
\usepackage{jheppub}
\usepackage{bm}
\usepackage{ulem}

\usepackage{pstricks}
\usepackage{color}

\def\I{{\cal I}}

\allowdisplaybreaks[3]

\begin{document}

\title{Composite operators in $T\bar T$-deformed free QFTs}

\author[]{Anshuman Dey,}
\author[]{Mikhail Goykhman,}
\author[]{Michael Smolkin}
\affiliation[]{The Racah Institute of Physics, The Hebrew University of Jerusalem, \\ Jerusalem 91904, Israel}
\emailAdd{anshuman.dey@mail.huji.ac.il}
\emailAdd{michael.goykhman@mail.huji.ac.il}
\emailAdd{michael.smolkin@mail.huji.ac.il}
%\emailAdd{}

\abstract{
We study perturbative renormalization of the composite operators in the $T\bar T$-deformed two-dimensional free field theories.  The  pattern of renormalization for the stress-energy tensor is different in the massive and massless cases. While in the latter case the canonical stress tensor is not renormalized up to high order in the perturbative expansion, in the massive theory there are induced counterterms at linear order.  For a massless theory our results match the general formula derived recently in \cite{Cardy:2019qao}. 
}

\maketitle

\section{Introduction}
\label{sec:introduction}

Recently, Smirnov and Zamolodchikov introduced a new class of tractable two-dimensional theories \cite{Smirnov:2016lqw}.  This class of models is rich. It is defined by the $T\bar T$ flow equation of the form
\begin{equation}
\label{DiffEI}
\frac{\partial \mathcal{L}(\lambda)}{\partial \lambda} = - 4 \left( T_{z z}^{\lambda}\,  T_{\bar z \bar z}^{\lambda} - (T_{z \bar z}^{\lambda})^2\right)~,
\end{equation}
where $\mathcal{L} (\lambda= 0)$ is any Lagrangian and $T_{\mu \nu}^{\lambda}$ are the components of the energy-momentum tensor of the finite $\lambda$ theory. The composite operators on both sides of this equation are UV finite. Although this equation has nothing to do with an RG flow equation, it does describe a particular one-parameter family of theories.  

Remarkably, the above prescription gives both the Lagrangian \cite{Cavaglia:2016oda, Bonelli:2018kik}, as well as the S-matrix \cite{Dubovsky:2012wk,Caselle:2013dra} and energy spectrum \cite{Zamolodchikov:2004ce,Dubovsky:2013ira} for these models. However, the S-matrices of the $T\bar T$ theories do not have a reliable analytic behavior -- they grow exponentially at large imaginary momenta. Such a growth is inconsistent with the behavior of a local quantum field theory. However, rather than just discarding these theories, the approach is to interpret them as quantum field theories coupled to gravity \cite{Dubovsky:2017cnj,Dubovsky:2018bmo,Cardy:2018sdv,Conti:2018tca,Conti:2018jho}. Hence, the $AdS_3$ dual of the $T\bar T$ deformed theories is an interesting question studied, for instance, in \cite{McGough:2016lol,Hartman:2018tkw,Taylor:2018xcy}, see also \cite{Giveon:2017nie,Chakraborty:2018vja,Giribet:2017imm,Nastase:2020evb,Guica:2020uhm,Barbon:2020amo} for recent developments on $T\bar T$ and string theory, and \cite{Gorbenko:2018oov} for studies of dS. Furthermore, at large $N$ one expects the $T\bar T$ deformation to represent a change in boundary conditions \cite{Guica:2019nzm}, whereas the Hagedorn behavior and partition function of the $T\bar T$ theories is an interesting question by itself \cite{Smirnov:2016lqw,Cavaglia:2016oda,Datta:2018thy,Jiang:2019hxb,Aharony:2018bad,Brennan:2020dkw}.

We consider low energy behavior of these theories. In this regime these are quantum field theories which can be studied perturbatively around $\lambda= 0$, see {\it e.g.,} \cite{Kraus:2018xrn,Rosenhaus:2019utc,He:2020udl,He:2020cxp,He:2020qcs}~\footnote{See also \cite{Haruna:2020wjw} where $T\bar T$ deformation was studied non-perturbatively in the large-$N$ limit of the $O(N)$ vector model.}. The renormalization of these theories is highly nontrivial. Indeed, they are non-renormalizable. If these were the standard QFTs, it would mean that one lacks predictive power in the UV. However, integrability gives an infinite number of constraints which uniquely fix all necessary counterterms, see {\it e.g.,} \cite{Rosenhaus:2019utc} for perturbative renormalization of the Lagrangian. In this work we focus on renormalization of the composite operators. Our approach is similar in spirit to the renormalization program in a conventional local renormalizable field theory, {\it i.e.,} we define deformed operators whose correlators are perturbatively finite. 

In the first part we demonstrate that, to
linear order in the $T\bar T$ deformation, renormalization of the composite (non-)primary operator ${\cal O}$
in a massless free field theory can be cast in the universal form
\begin{equation}
\label{universal renormalization}
[{\cal O}] = {\cal O} - \frac{\lambda}{2\pi\epsilon}\,\partial^2\,{\cal O}\,,
\end{equation}
where $\epsilon=2-d$ regulates a logarithmic UV divergence, whereas ${\cal O}$ and $[{\cal O}]$ represent the bare and renormalized operators respectively.
The general argument in favour of this universal relation
was recently provided in \cite{Cardy:2019qao}.
We explicitly verify it using the examples of scalar and spinning/tensor (non-)primary operators ${\cal O}$ in the case of massless Dirac and scalar free field theories deformed by the $T\bar T$ operator.
In particular, we study renormalization of the composite operators such as $\phi^n$, $\left((\partial\phi)^2\right)^n$,
$(\bar\psi\psi)^n$ (for any integer $n\geq 1$), $\partial_\mu\phi\partial_\nu\phi$, $\bar\psi\gamma^\mu\psi$, $\bar\psi\gamma_\mu\partial_\nu\psi$,
and the stress-energy tensor $T_{\mu\nu}$. 
The renormalization pattern eventually matches (\ref{universal renormalization}) for all cases.

We argue that the canonical stress-energy tensor for the massless Dirac
and scalar free field theories is not renormalized to linear order in the coupling $\lambda$.
This contrasts significantly with the $\phi^4$ theory in $d=4-\epsilon$
dimensions, where it is necessary to introduce a non-trivial improvement term 
$\frac{g}{\epsilon}(\partial_\mu\partial_\nu-\delta_{\mu\nu}\partial^2)\phi^2$ to render the canonical stress-energy tensor finite \cite{Callan:1970ze}.\footnote{Here $g$ is the $\phi^4$ coupling constant.} Since improvement terms can be attributed to the corresponding gravitational counterterms, {\it e.g.,} $\frac{g}{\epsilon}R\phi^2$ in the case of the $\phi^4$ theory \cite{Collins:1965}, they partially elucidate the way in which a quantum field theory couples to a curved background. This is particularly interesting in the context of $T\bar T$ theories since little is known about how such theories couple to a curved manifold.
%is to be contrasted with appearance of a  non-trivial improvement term
%$\frac{g}{\epsilon}(\partial_\mu\partial_\nu-\delta_{\mu\nu}\partial^2)\phi^2$ which is required
%to renormalize the composite stress-energy tensor operator in the $\phi^4$ theory in $d=4-\epsilon$
%dimensions,
%and which can be attributed to the corresponding gravitational counterterm $\frac{g}{\epsilon}R\phi^2$
%where $g$ is the quartic coupling \cite{Collins:1965}.

Note that the Ward identity for the correlation functions involving a divergence of the stress-energy 
tensor (or any other conserved current) implies a non-renormalization of this operator.\footnote{See \cite{Kraus:2018xrn} for a related discussion in the $T\bar T$ context.} However, the potentially singular in the $\epsilon\rightarrow 0$
limit improvement terms, which contribute to the stress-energy
tensor, are not in conflict with this satement, because
they are separately conserved.

%We notice that the $T\bar T$-deformed theories of massless scalar and fermion in $d=2-\epsilon$ dimensions
%in general also admit identically conserved improvement counterterms contributing to the renormalized
 %stress-energy tensor.
 
In fact, one can list the admissible improvement counterterms. To this end, we recall that such (divergent) terms are local, and can be derived by varying the induced gravitational counterterms with respect to the metric around a flat background. While the coupling of $T\bar T$ theories to gravity remains a challenge, there is no obstruction to  present an exhaustive list of possible counterterms based on dimensional analysis. 
 %While the coupling of $T\bar T$ theories to gravity remains a challenge, we can still formally and
%concisely formulate such improvement counterterms as originating from varying the
%gravitational counterterms w.r.t. the metric, and subsequently setting the metric to be flat.
%The corresponding gravitational counterterms 
For instance, to linear order in the $T\bar T$ deformation, we have the following candidates for the massless scalar and Dirac fields
\begin{align}
\label{general grav ct for scalar}
I^{\textrm{g.c.t}}_{(\textrm{s})} &= \frac{\lambda}{\epsilon}\,\int d^dx\,
\left(b^{(s)}_1 R_{\mu\nu}+b^{(s)}_2  R g_{\mu\nu}\right)\,\partial^\mu\phi\partial^\nu\phi\,,\\
\label{general grav ct for fermion}
I^{\textrm{g.c.t}}_{(\textrm{f})} &= \frac{\lambda}{\epsilon}\,\int d^dx\,
\left(b^{(f)}_1 R_{\mu\nu}+b^{(f)}_2  R g_{\mu\nu}\right)\,\bar\psi\gamma^\mu\partial^\nu\psi\,.
\end{align}
In what follows we show that the coefficients $b^{(s,f)}_{1,2}$ for the free fields satisfy
\begin{equation}
b^{(s,f)}_2=-\frac{1}{2}b^{(s,f)}_1\,.
\end{equation}
Hence, the right hand side of (\ref{general grav ct for scalar}) and (\ref{general grav ct for fermion}) is proportional to the Einstein tensor,
which identically vanishes in two dimensions. 
%In subsection \ref{subsec: partial mu phi partial nu phi} we illustrate
%how the use of gravitational counterterm (\ref{general grav ct for scalar}) makes
%it much more transparent to see why the associated and cumbersome-looking
%improvement contribution to the stress-energy tensor is identically conserved.
Note that the non-renormalization of the stress-energy tensor in the $T\bar T$-deformed free theories is in full
agreement with (\ref{universal renormalization}), because $\partial^2T_{\mu\nu}$ vanishes in a CFT.\footnote{We
are using equations of motion to the zeroth order in the deformation, because (\ref{universal renormalization}) holds to linear order in the coupling $\lambda$.}

Finally, in section \ref{sec: massive case} we study composite operators in the massive free field theories. The renormalization pattern does not match (\ref{universal renormalization}) in this case.  Thus, for instance, to linear order in the deformation there are induced counterterms in the stress-energy tensor which do not obey a simple relation (\ref{universal renormalization}). This suggests that unlike the case of deformed CFT, presence of a gap in the undeformed theory results in an operator mixing under RG flow for any non-zero $\lambda$. It would be interesting to explore its structure in the future.

\section{Renormalization of composite operators: scalar field}
\label{sec:scalar}

In this section we consider $T\bar T$ deformation of the free massless scalar field. 
To the linear order in the  $T\bar T$ coupling $\lambda$ the action is given by
\begin{align}
I=\int d^dx\, \left(\frac{1}{2}\, \partial_\mu \phi \partial ^\mu\phi
-\frac{\lambda\, \mu^{\epsilon}}{4}\, \left((\partial  \phi)^2\right)^2\right)\,,
\label{renormalized action}
\end{align}
where $\mu$ is an arbitrary renormalization scale, and
the leading order $T\bar T$ perturbation is given by
\begin{equation}
\label{TT0 in terms of phi}
T\bar T_0 \equiv \det T^{(0)} = -\frac{1}{4}\,\left((\partial  \phi)^2\right)^2\,.
\end{equation}
Here label $0$ indicates zeroth order in the coupling $\lambda$,
and we work in dimensional regularization, $d=2-\epsilon$.
We will perform most of our calculations to the linear order in the $T\bar T$
coupling $\lambda$, in which case all the counterterms in the action (\ref{renormalized action}) vanish,
and in particular $\lambda$ is the renormalized dimensionless coupling, and the bare fields 
coincide with the renormalized fields.

Stress-energy tensor to linear order in $\lambda$ is given by
\begin{align}
\label{Tmunu scalar full expression}
T_{\mu\nu} &= \partial_\mu\phi \partial_\nu\phi  (1-\lambda(\partial\phi)^2)-
\frac{1}{2}\,\delta_{\mu\nu}\,\left((\partial\phi)^2+2\,\lambda\, T\bar T\right)\,,
\end{align}
To avoid clutter in our notation, here and in what follows we will skip putting a label on the stress-energy
tensor to indicate to which order in $\lambda$ it is written.

Free massless scalar two-point function is given by
\begin{equation}
\langle \phi(x)\phi(0)\rangle  = \frac{\Gamma\left(\frac{d-2}{2}\right)}{4\pi^\frac{d}{2}}\,\frac{1}{|x|^{d-2}}\,.
\end{equation}
In all of our calculations we are looking
for logarithmically divergent terms, which (to linear order in the coupling $\lambda$)
 in dimensional regularization correspond to only simple
poles in $\epsilon=2-d$. Therefore to simplify some of our notations we therefore can and will set $d=2$
($\epsilon=0$)
in some of the factors right away.

\subsection{$[\phi^n]$}
Consider renormalization of the composite operator $\phi^n$, $n\geq 2$
in the theory of massless scalar field $\phi$ to the linear order in the $T\bar T$
coupling $\lambda$,
\begin{equation}
\phi^n = [\phi^n] + \Delta \phi ^ n \,, \qquad
\Delta \phi ^ n =  \frac{\lambda\,\mu^\epsilon}{4}\,
\phi ^ n \, \int d^dx \, \left( (\partial\phi)^2 \right) ^ 2\,.
\end{equation}
where in $\Delta \phi ^ n$ we implicitly assumed to perform all possible contractions and keep only terms singular 
in the $d\rightarrow 2$ limit.
In this section, just as everywhere else in this paper, we are working in the dimensional regularization, and focus on
contributions to $\Delta \phi ^ n$ which are divergent in the $d\rightarrow 2$ limit.
For simplicity and without loss of generality we also consider the operator $\phi^n$ (and all other composite operators
studied in this paper) to sit at the origin $x=0$.

We begin by writing possible contractions which can contribute to the divergent terms in the $d\rightarrow 2$ limit,
\footnote{Here the factor of $\frac{n(n-1)}{2!}$ is the combinatorial coefficient due to
the choice of two out of $n$ operators $\phi(0)$, and the factor of $4$ is due to the possible choices
of $\partial^\nu\phi$ in the first line of (\ref{Delta phi  n}).}
\begin{align}
\Delta \phi ^ n &= \frac{\lambda\,\mu^\epsilon}{4}\, \frac{n(n-1)}{2!} \, 4 \, \phi^{n-2}\,
\int d^d x \, \langle \phi(0)\partial^\nu \phi (x) \rangle \label{Delta phi  n}\\
&\times\left( \langle \phi (0)\partial_\nu \phi (x)\rangle (\partial\phi(x)) ^ 2
+ 2 \partial_\nu \phi (x)\partial _\mu \phi (x) \langle \phi (0) \partial^\mu \phi (x)\rangle \right)\,.
\label{Delta phi ^ n preliminary}
\end{align}
As mentioned above,
to the linear order in the coupling $\lambda$ we will only encounter at most simple poles in $1/\epsilon$, where
$\epsilon = 2 - d$. Therefore we can set $d=2$ everywhere else, and in particular we substitute here
\begin{equation}
\label{first derivative of phi massless propagator}
\langle \phi(0) \partial^\nu \phi(x) \rangle = - \frac{1}{2\pi}\,\frac{x^\nu}{|x|^d}\,.
\end{equation}
Since we are interested in the UV divergencies only, we will focus on the structure
of the integral (\ref{Delta phi ^ n preliminary}) near the origin $x=0$. Expanding the integrated operators
around $x=0$,
and keeping only the terms logarithmically divergent in $d=2$ (equivalently,
keeping only the terms which exhibit $1/\epsilon$ poles in the $d\rightarrow 2$ limit),
we obtain
\begin{equation}
\Delta \phi ^ n  = \frac{\lambda n (n-1)}{8\pi^2} \, \phi^{n-2} \,
\left({\cal I}\, (\partial \phi)^2 + 2 {\cal I}^{\mu\nu} \, \partial_\mu \phi \partial_\nu\phi\right)\,,
\end{equation}
where ${\cal I}$, ${\cal I}^{\mu\nu}$ are defined in (\ref{master integral general definition}).
Using (\ref{I0 expression}), (\ref{I2 expression}) we obtain
\begin{equation}
\label{Delta phi to n before eom}
\Delta \phi ^ n = \frac{\lambda n (n-1)}{2\pi\epsilon} \, \phi^{n-2} \, (\partial\phi)^2 \,.
\end{equation}

Using equations of motion for the field $\phi$ to the ${\cal O}(\lambda^0)$ order, $\partial^2\phi=0$,
we can re-write the order ${\cal O}(\lambda)$ expression on the r.h.s. of (\ref{Delta phi to n before eom}) as
\begin{equation}
\label{Delta phi to n after eom}
\Delta \phi ^ n = \frac{\lambda}{2\pi\epsilon} \, \partial^2 \phi^{n} + {\cal O}(\lambda^2) \,.
\end{equation}
The expression is in agreement with the general form (\ref{universal renormalization})
stated in Introduction.

\subsection{$[\partial_\mu\phi\partial_\nu\phi]$}
\label{subsec: partial mu phi partial nu phi}
We next  proceed to calculate renormalization of the composite tensor operator $\partial_\mu\phi\partial_\nu\phi$,
\begin{align}
\label{d mu phi d nu phi start}
\partial_\mu\phi\partial_\nu\phi &= [\partial_\mu\phi\partial_\nu\phi] + \Delta \partial_\mu\phi\partial_\nu\phi\,, \\
\Delta \partial_\mu\phi\partial_\nu\phi &= \frac{\lambda\,\mu^\epsilon}{4}\,
\int d^dx \,\left( 4\, \partial_\mu\partial_\alpha \langle \phi(0)\phi(x)\rangle
\partial_\nu\partial^\alpha \langle \phi(0)\phi(x)\rangle (\partial\phi(x))^2\right.\\
&+\left. 8\, \partial_\mu\partial_\alpha \langle \phi(0)\phi(x)\rangle
\partial_\nu\partial_\beta \langle \phi(0)\phi(x)\rangle \partial^\alpha\phi(x)\partial^\beta\phi(x) \right)\,,
\end{align}
where we have kept track of the various degeneracy factors originating from combinatorics.
We are ultimately interested in collecting the terms which are singular in the $\epsilon\rightarrow 0$ limit.
To this end, using the free correlation function
\begin{align}
\langle \phi(0) \partial_\mu\partial_\alpha\phi(x)\rangle = \frac{1}{\pi}
\frac{1}{|x|^{d+2}}\,\left(x_\alpha x_\mu -\frac{\delta_{\alpha\mu}}{2}\,|x|^2\right)
\end{align}
and
expanding the integrated operators around $x=0$ we obtain
\begin{align}
\label{Delta d mu phi d nu phi intermediate result preliminary}
\Delta \partial_\mu\phi\partial_\nu\phi  &=\frac{\lambda}{2\pi^2}\,
\left[\frac{1}{4}\,\delta_{\mu\nu}\,\I_{\alpha\beta}\,\partial^\alpha\partial^\beta ((\partial \phi(0))^2 )\right.\\
&+\left. 2\, \partial^\rho\partial^\sigma (\partial^\alpha\phi
\partial^\beta\phi(0))\left(I_{\rho\sigma\alpha\mu\beta\nu} -\frac{1}{2}\,
\delta_{\beta\nu}\, \I_{\rho\sigma\alpha\mu}-
\frac{1}{2}\, \delta_{\alpha\mu} \, \I_{\rho\sigma\beta\nu}
+\frac{1}{4}\, \I_{\rho\sigma}\, \delta_{\alpha\mu}\, \delta_{\beta\nu}\right)\right]\,.\notag
\end{align}
This expression can be further simplified using (\ref{I2 expression}), (\ref{I4 expression}), (\ref{I6 expression}).
While the calculation is rather tedious, with some insight it can be automatized with the help of \textit{Mathematica}, rendering 
\begin{align}
\label{Delta d mu phi d nu phi intermediate result}
\Delta \partial_\mu\phi\partial_\nu\phi &=\frac{\lambda}{12\pi\epsilon}
\left[\left(2\partial^2(\partial\phi)^2+\partial_\alpha\partial_\beta (\partial^\alpha\phi
\partial^\beta\phi)\right)\delta_{\mu\nu}
+\partial_\mu\partial_\nu (\partial\phi)^2 +\partial^2 (\partial_\mu\phi\partial_\nu\phi)\right.\notag\\
&-\left.\partial_\alpha\partial_\mu (\partial^\alpha\phi\partial_\nu\phi)
-\partial_\alpha\partial_\nu (\partial^\alpha\phi\partial_\mu\phi)\right]\,.
\end{align}

As a quick detour, notice that from (\ref{Delta d mu phi d nu phi intermediate result}) we can immediately
 deduce renormalization
of the composite scalar operator $(\partial\phi)^2$,
\begin{equation}
\label{partial phi squared renormalization}
(\partial \phi)^2 = [(\partial \phi)^2] +\frac{\lambda}{2\pi\epsilon}\,\partial^2(\partial\phi)^2 + {\cal O}(\lambda^2)\,,
\end{equation}
which is in agreement with the general form (\ref{universal renormalization})
stated in Introduction. Below we will generalize this result to renormalization of $\left((\partial\phi)^2\right)^n$
for arbitrary $n\geq 1$.

Returning back to (\ref{Delta d mu phi d nu phi intermediate result}) we notice that it can be further
simplified using tensor identities in $d=2$. Specifically, using the following variations around the flat metric
\begin{align}
\notag
\int d^d x\, (\partial\phi)^2\, g^{\mu\nu}\,\delta R_{\mu\nu}
&=\int d^2 x\, (\partial\phi)^2\, (g_{\mu\nu}\,\nabla^2 - \nabla_\mu\nabla_\nu)\delta g^{\mu\nu}= \int d^2x\, \delta g^{\mu\nu}\,(\partial^2\delta_{\mu\nu}-\partial_\mu\partial_\nu) (\partial\phi)^2\,,\\
\int d^2x \, \partial ^ \alpha \phi \partial ^ \beta\phi\,  \delta R_{\alpha\beta}
&=\frac{1}{2}\int d^2x  \delta g^{\mu\nu}
(\partial_\alpha\partial_\beta (\partial^\alpha\phi\partial^\beta\phi) \delta_{\mu\nu}
{+}\partial^2 (\partial_\mu\phi\partial_\nu\phi) 
{-} \partial_\mu\partial_\alpha(\partial_\nu\phi\partial^\alpha\phi)
{-}\partial_\nu\partial_\alpha(\partial_\mu\phi\partial^\alpha\phi))\,,\notag
\end{align}
we obtain \footnote{It is possible to verify directly that the l.h.s. of (\ref{2d einstein identity})
vanishes in $2d$, particularly by plugging that expression into \textit{Mathematica}.}
\begin{align}
&\left(-\partial^2(\partial\phi)^2
+\partial_\alpha\partial_\beta(\partial^\alpha\phi\partial^\beta\phi)
\right)\,\delta_{\mu\nu}
+\partial_\mu\partial_\nu (\partial\phi)^2
+\partial^2(\partial_\mu\phi\partial_\nu\phi)
-\partial_\alpha\partial_\mu(\partial^\alpha\phi\partial_\nu\phi)
-\partial_\alpha\partial_\nu(\partial^\alpha\phi\partial_\mu\phi)\notag\\
&=\frac{\delta}{\delta g^{\mu\nu}}\int  d^2x\,
(-R \, g_{\alpha\beta} +2R_{\alpha\beta})\,\partial^\alpha\phi\partial^\beta\phi \equiv 0\,,
\label{2d einstein identity}
\end{align}
where in the last line we took advantage of the fact that the Einstein tensor $R_{\mu\nu}-\frac{1}{2}R\,g_{\mu\nu}$
vanishes identically in two dimensions.

Using (\ref{2d einstein identity}) we can re-write (\ref{Delta d mu phi d nu phi intermediate result}) as
\begin{equation}
\label{Delta d mu phi d nu phi simplified result}
\Delta \partial_\mu\phi\partial_\nu\phi =\frac{\lambda}{4\pi\epsilon}\,
\partial^2(\partial\phi)^2\, \delta_{\mu\nu}\,.
\end{equation}
Finally, using the ${\cal O}(\lambda^0)$ order equations of motion $\partial^2\phi = 0$, one can show
that in $d=2$ the ${\cal O}(\lambda)$ expression (\ref{Delta d mu phi d nu phi simplified result}) is equivalent to
\begin{equation}
\label{Delta d mu phi d nu phi final result}
\Delta \partial_\mu\phi\partial_\nu\phi =\frac{\lambda}{2\pi\epsilon}\,
\partial^2(\partial_\mu\phi\partial_\nu\phi)\,,
\end{equation}
in agreement with (\ref{universal renormalization}).

\subsection{$[T_{\mu\nu}]$}

Above we have accumulated enough results to calculate renormalization of the stress-energy
tensor (\ref{Tmunu scalar full expression}),
\begin{equation}
T_{\mu\nu}= [T_{\mu\nu}] + \Delta T_{\mu\nu}
\end{equation}
to linear order in the $T\bar T$ couping $\lambda$.
First of all we notice that in massless theory the term $\partial_\mu\phi \partial_\nu\phi (\partial\phi)^2$
does not get renormalized to zeroth order in $\lambda$. Then, using 
(\ref{Delta d mu phi d nu phi simplified result}) we obtain (without using equations of motion for the field $\phi$)
\begin{equation}
\Delta T_{\mu\nu} = 0 + {\cal O}(\lambda^2)\,.
\end{equation}
Notice that while this simple result is in an apparent conflict with the general formula (\ref{universal renormalization}),
the disagreement is superficial. Indeed, as we have argued above, the $\partial_\mu\phi\partial_\nu\phi$
is renormalized according to (\ref{universal renormalization}) once the equations of motion are used, see discussion
leading to (\ref{Delta d mu phi d nu phi final result}). Since to linear order in $\lambda$
the renormalization of $\partial_\mu\phi\partial_\nu\phi$ defines renormalization of $T_{\mu\nu}$, we conclude that the ${\cal O}(\lambda^0)$ equations of motion imply
\begin{equation}
\label{vanishing d squared for scalar T}
\partial^2 T_{\mu\nu} = 0\,,
\end{equation}
which holds in a CFT.
%The expression (\ref{vanishing d squared for scalar T})
%can of course be explicitly verified in $2d$, in particular with the aid of \textit{Mathematica}.

%As advertised in Introduction, due to the general argument in \cite{Cardy:2019qao}
%which supplies us with exact evolution equation for any composite operator, we conclude that
%\begin{equation}
%[T_{\mu\nu}] = T_{\mu\nu}
%\end{equation}
%to all orders in coupling $\lambda$.

\subsection{$[((\partial\phi)^2)^n]$}
\label{subsec:partial phi n}

To conclude our discussion of the $T\bar T$-deformed free massless scalar we 
generalize our result (\ref{partial phi squared renormalization}) to the case of arbitrary $n\geq 1$
\begin{equation}
\left((\partial \phi)^2\right)^n = [\left((\partial \phi)^2\right)^n] +\Delta \left((\partial \phi)^2\right)^n\,.
\end{equation}
Here we have\footnote{Notice that we contract at most two pairs of fields. It is possible
to contract one or two more fields for $n\geq 3$, but such extra terms will not contribute any
logarithmic divergencies.}
\begin{align}
\Delta \left((\partial \phi)^2\right)^n &= \frac{\lambda}{4}\,
\left((\partial \phi)^2\right)^n\,\int d^dx\, \left((\partial\phi)^2\right)^2\\
&=n\,\left((\partial \phi)^2\right)^{n-1}\,\Delta (\partial \phi)^2
+\lambda \, n\, (n-1)\, \left((\partial \phi)^2\right)^{n-2}\,\partial_\mu\phi \partial^\alpha\phi\notag \\
&\times \int d^dx\, \langle \partial^\mu\phi(0)\partial^\nu\phi(x)\rangle \,
\left((\partial\phi(x))^2\,\langle \partial_\alpha\phi(0)\partial_\nu\phi(x)\rangle 
+2\,\langle\partial_\alpha\phi(0)\partial_\beta\phi(x)\rangle\partial^\beta\phi(x)\partial_\nu\phi(x)\right)\,.\notag
\end{align}
Expanding the integrated operator around $x=0$ and keeping only the logarithmically divergent
contributions we obtain
\begin{align}
\Delta \left((\partial \phi)^2\right)^n &=n\,\left((\partial \phi)^2\right)^{n-1}\,\Delta (\partial \phi)^2
+\frac{\lambda n (n-1)}{2\pi^2}\,((\partial\phi)^2)^{n-2}\,
\left(\frac{1}{2}\,(\partial\phi)^2\,\I^{\alpha\beta}\,\partial_\alpha\partial_\beta ((\partial\phi)^2)\right.\\
&+\left. 4\partial_\mu\phi \partial^\alpha\phi \left(\I^{\mu\nu\lambda\rho}_{\;\;\;\;\;\;\;\;\alpha\beta}-
\frac{1}{2}\I^{\mu\nu\lambda\rho}\delta_{\alpha\beta}-\frac{1}{2}\I^{\lambda\rho}_{\;\;\;\;\alpha\beta}
\delta^{\mu\nu}+\frac{1}{4}\I^{\lambda\rho}\delta^{\mu\nu}\delta_{\alpha\beta}\right)\right)
\partial_\lambda\partial_\rho(\partial^\beta\phi\partial_\nu\phi)\,.\notag
\end{align}
Plugging here (\ref{I2 expression}), (\ref{I4 expression}),  (\ref{I6 expression}), and simplifying
it in \textit{Mathematica}, we obtain
\begin{align}
\label{Delta d phi squared to n preliminary result}
\Delta \left((\partial \phi)^2\right)^n =n\,\left((\partial \phi)^2\right)^{n-1}\,\Delta (\partial \phi)^2
+\frac{\lambda n(n-1)}{2\pi\epsilon}\,\left((\partial \phi)^2\right)^{n-1}\,
\partial^2\left((\partial\phi)^2\right)\,.
\end{align}
Using (\ref{partial phi squared renormalization}) we can re-write it as
\begin{equation}
\label{Delta d phi squared to n preliminary 2 result}
\Delta \left((\partial \phi)^2\right)^n =
\frac{\lambda n^2}{2\pi\epsilon}\,\left((\partial \phi)^2\right)^{n-1}\,
\partial^2\left((\partial\phi)^2\right)\,.
\end{equation}

On the other hand,
\begin{align}
\partial^2 \left((\partial\phi)^2\right)^n &=
n\,\left((\partial\phi)^2\right)^{n-2}\,\left((\partial\phi)^2\,\partial^2\,(\partial\phi)^2
+(n-1)\,\partial^\mu(\partial\phi)^2\,\partial_\mu(\partial\phi)^2\right)\,.
\end{align}
Using equations of motion $\partial^2\phi = 0$ we can demonstrate that in two dimensions
$\partial^\mu(\partial\phi)^2\,\partial_\mu(\partial\phi)^2 = (\partial\phi)^2\,\partial^2\,(\partial\phi)^2$.
Consequently
\begin{align}
\partial^2 \left((\partial\phi)^2\right)^n &=
n^2\,\left((\partial\phi)^2\right)^{n-1}\,\partial^2\,(\partial\phi)^2\,.
\end{align}
Comparing this with (\ref{Delta d phi squared to n preliminary 2 result}) we conclude
\begin{equation}
\label{Delta d phi squared to n final result}
\Delta \left((\partial \phi)^2\right)^n =
\frac{\lambda }{2\pi\epsilon}\, \partial^2 \left((\partial\phi)^2\right)^n\,,
\end{equation}
in agreement with (\ref{universal renormalization}).

\section{Renormalization of composite operators: Dirac field}
\label{sec:fermion}

In this section we will recreate analysis performed in section \ref{sec:scalar} for the theory of free
massless Dirac fermion deformed by the $T\bar T$ operator.
We will be working to the linear order in the $T\bar T$ coupling $\lambda$. 
The corresponding Euclidean action
to linear order in  $\lambda$ is given by
\begin{equation}
I =\int d^dx\,\left(
 \frac{1}{2}\, (\bar\psi \gamma^\mu \partial _\mu \psi - \partial_\mu\bar\psi \gamma^\mu \psi)
-\frac{\lambda\,\mu^\epsilon}{2}\,T_{\mu\nu}T^{\mu\nu} 
 \right)\,.
\end{equation}
Here we have used the fact that in two dimensions for traceless stress-energy tensor one can write down
\begin{equation}
\det T = -\frac{1}{2}\, T_{\mu\nu}T^{\mu\nu} \,.
\end{equation}
The canonical stress-energy tensor for the free fermion is given by
\begin{equation}
T_{\mu\nu}^{(c)} = \frac{1}{2}\,(\bar\psi \gamma_\mu\partial_\nu \psi - \partial_\nu\bar\psi\gamma_\mu\psi)\,.
\end{equation}
Using the standard Belinfante technique we can symmetrize it, giving
\begin{equation}
\label{fermion stress energy tensor}
T_{\mu\nu} = \frac{1}{4}\,(\bar\psi \gamma_\mu\partial_\nu \psi
+\bar\psi \gamma_\nu\partial_\mu \psi - \partial_\nu\bar\psi\gamma_\mu\psi
 - \partial_\mu\bar\psi\gamma_\nu\psi)\,.
\end{equation}

Free massless fermion two-point function is given by
\begin{equation}
\langle \psi(x)\bar\psi(0)\rangle  = \frac{\Gamma\left(\frac{d}{2}\right)}{2\pi^\frac{d}{2}}\,\frac{x^\mu\gamma_\mu}
{|x|^{d}}\,.
\end{equation}

\subsection{$[(\bar\psi\psi)^n]$}
\label{subsec: bar psi psi}

In this subsection we consider one-loop renormalization of the composite operator $(\bar\psi\psi)^n$
for $n\geq 1$. Let us begin by studying the $n=1$ case,
\begin{align}
\bar\psi\psi = [\bar\psi\psi] + \Delta\bar\psi\psi\,,\quad
\Delta\bar\psi\psi = \frac{\lambda\,\mu^\epsilon}{2}\,\bar\psi\psi \, \int d^dx\, T_{\lambda\rho}T^{\lambda\rho}\,,
\end{align}
where again we imply performing all possible contractions on the r.h.s. of the
expression for $\Delta\bar\psi\psi$,
and keeping only singular terms in the
 $d\rightarrow 2$ limit.
 %As we did when calculating $[\bar\psi\gamma_\mu\partial_\nu\psi]$
It is convenient to split $\Delta\bar\psi\psi$ into two contributions,
\begin{equation}
\Delta\bar\psi\psi = \Delta _{(1)}\bar\psi\psi + \Delta _{(2)}\bar\psi\psi\,,
\end{equation}
where the first contribution is due to 
contractions of constituent fermions in $\bar\psi\psi$ with any of the two factors $T_{\lambda\rho}$
in the vertex $T_{\lambda\rho}T^{\lambda\rho}$. Taking into account symmetry of stress-energy tensor and
real-value-ness of several equal to each other terms, we write down
\begin{align}
\Delta _{(1)}\bar\psi\psi &= \frac{\lambda}{4}\,\int d^dx\,
\textrm{Tr}\,\left(\langle \psi(0)\bar\psi(x)\rangle \,\gamma_\lambda\,
\langle \partial_\rho\psi(x)\bar\psi(0)\rangle\right) T^{\lambda\rho}\\
&=\frac{\lambda}{16\pi^2}\left(\I^{\mu\alpha}\,\textrm{Tr}\,(\gamma_\mu\gamma_\lambda
\gamma_\rho)-2\I^{\mu\nu\alpha}_{\quad\;\;\;\rho}\,\textrm{Tr}\,
(\gamma_\mu\gamma_\lambda\gamma_\nu)\right)\,\partial_\alpha \,T^{\lambda\rho}\,.
\end{align}
One can verify explicitly that each individual term on the r.h.s. of the last line vanishes.

The second contribution to $\Delta\bar\psi\psi$ originates from cross-contraction between
constituent fermions of $\bar\psi\psi$ and each of the two stress-energy factors in the
vertex $T_{\lambda\rho}T^{\lambda\rho}$. Taking into account all possible degeneracy factors, we write down
\begin{align}
\notag
\Delta _{(2)}&\bar\psi\psi {=}\frac{\lambda}{16} \left(
2\int d^dx\left(\bar\psi\gamma^\lambda\langle \partial^\rho\psi(x)\bar\psi(0)\rangle
\langle \psi(0)\bar\psi(x)\rangle \gamma_\lambda\partial_\rho\psi
{+}\bar\psi\gamma^\lambda\langle \partial^\rho\psi(x)\bar\psi(0)\rangle
\langle \psi(0)\bar\psi(x)\rangle \gamma_\rho\partial_\lambda\psi\right)
\right.\\
&-\left.\int d^dx\left(
\partial^\lambda\bar\psi\gamma^\rho\langle \psi(x)\bar\psi(0)\rangle
\langle \psi(0)\bar\psi(x)\rangle \gamma_\lambda\partial_\rho\psi+
\partial^\rho\bar\psi\gamma^\lambda\langle \psi(x)\bar\psi(0)\rangle
\langle \psi(0)\bar\psi(x)\rangle \gamma_\lambda\partial_\rho\psi
\right)
\right.\notag\\
&-\left.
\int d^dx\left(
\bar\psi\gamma_\lambda\langle \partial_\rho\psi(x)\bar\psi(0)\rangle
\langle \psi(0)\partial^\lambda\bar\psi(x)\rangle \gamma^\rho\psi
+\bar\psi\gamma_\rho\langle \partial_\lambda\psi(x)\bar\psi(0)\rangle
\langle \psi(0)\partial^\lambda\bar\psi(x)\rangle \gamma^\rho\psi
\right)+\textrm{c.c.}
\right)\,.
\end{align}
To explain the calculation we split this into the first, second, and third line contributions
(each of them plus their complex conjugate contributions,
where we find out that the second and third lines are actually individually real-valued),
\begin{equation}
\Delta _{(2)}\bar\psi\psi = \Delta _{(2)}^{(1)}\bar\psi\psi 
+\Delta _{(2)}^{(2)}\bar\psi\psi
+\Delta _{(2)}^{(3)}\bar\psi\psi\,,
\end{equation}
where
\begin{align}
\Delta _{(2)}^{(1)}\bar\psi\psi&=\frac{\lambda}{64\pi^2}\,
\left(
-2\partial_\alpha(\bar\psi\gamma^\lambda\gamma^\rho\gamma_\nu\gamma_\lambda\partial_\rho\psi)\I^{\alpha\nu}
+4\partial_\alpha(\bar\psi\gamma^\lambda\gamma_\mu\gamma_\nu\gamma_\lambda\partial_\rho\psi)
\I^{\alpha\mu\nu\rho}\right.\\
&\left.-2\partial_\alpha(\bar\psi\gamma^\lambda\gamma^\rho\gamma_\nu\gamma_\rho\partial_\lambda\psi)\I^{\alpha\nu}
+4\partial_\alpha(\bar\psi\gamma^\lambda\gamma_\mu\gamma_\nu\gamma_\rho\partial_\lambda\psi)
\I^{\alpha\mu\nu\rho}
\right) +\textrm{c.c.}\,\notag
\end{align}
Using here (\ref{I2 expression}), (\ref{I4 expression}) we can simplify it to
\begin{equation}
\Delta _{(2)}^{(1)}\bar\psi\psi
=\frac{\lambda}{8\pi\epsilon}\,
\left(
\partial^2(\bar\psi\psi)-2(\partial_\mu\bar\psi\gamma^\mu\gamma^\nu\partial_\nu\psi-\partial_\mu
\bar\psi\partial^\mu\psi)
\right)\,.
\end{equation}
Further using ${\cal O}(\lambda^0)$ e.o.m in this ${\cal O}(\lambda)$ expression we can re-write it as
\begin{equation}
\Delta _{(2)}^{(1)}\bar\psi\psi = \frac{\lambda}{4\pi\epsilon}\,\partial^2(\bar\psi\psi)\,.
\end{equation}
Next, we have
\begin{align}
\Delta _{(2)}^{(2)}\bar\psi\psi&=\frac{\lambda}{32\pi^2}\left(
\partial^\lambda\bar\psi\gamma^\rho\gamma_\lambda\partial_\rho\psi
+2\partial^\lambda\bar\psi\partial_\lambda\psi
\right)\,\I = \frac{\lambda}{8\pi\epsilon}\,\partial^2(\bar\psi\psi)\,,
\end{align}
where in the last line we used (\ref{I0 expression}) and ${\cal O}(\lambda^0)$ equations of motion. 
Finally, 
\begin{align}
\Delta _{(2)}^{(3)}\bar\psi\psi&=\frac{\lambda}{32\pi^2}\,
\left(
\frac{1}{2}\partial_\alpha\partial_\beta (\bar\psi\gamma_\lambda\gamma_\rho\gamma^\lambda\gamma^\rho\psi)
\,\I^{\alpha\beta} -\partial_\alpha\partial_\beta (\bar\psi\gamma_\lambda\gamma_\rho\gamma_\nu\gamma^\rho\psi)
\,\I^{\alpha\beta\lambda\nu}\right.\\
&-\left.\partial_\alpha\partial_\beta (\bar\psi\gamma_\lambda\gamma_\nu\gamma^\lambda\gamma_\rho\psi)
\,\I^{\alpha\beta\nu\rho}
+2\partial_\alpha\partial_\beta (\bar\psi\gamma_\lambda\gamma_\rho\psi)\,\I^{\alpha\beta\lambda\rho}
\right.\notag\\
&+\left.\frac{1}{2}\partial_\alpha\partial_\beta (\bar\psi\gamma_\lambda\gamma_\rho\gamma^\rho\gamma^\lambda\psi)
\,\I^{\alpha\beta} -\partial_\alpha\partial_\beta (\bar\psi\gamma_\lambda\gamma_\rho\gamma_\nu\gamma^\lambda\psi)
\,\I^{\alpha\beta\rho\nu}\right.\notag\\
&-\left.\partial_\alpha\partial_\beta (\bar\psi\gamma_\lambda\gamma_\nu\gamma_\rho\gamma^\lambda\psi)
\,\I^{\alpha\beta\nu\rho}
+2\partial_\alpha\partial_\beta (\bar\psi\gamma_\lambda\gamma^\lambda\psi)\,\I^{\alpha\beta\rho}_{\quad\;\;\rho}
\right)\,.\notag
\end{align}
Using here (\ref{I2 expression}), (\ref{I4 expression}) only, we can simplify it to
\begin{equation}
\Delta _{(2)}^{(3)}\bar\psi\psi=\frac{\lambda}{8\pi\epsilon}\,\partial^2(\bar\psi\psi)\,.
\end{equation}
Combining everything together we obtain
\begin{equation}
\Delta _{(2)}\bar\psi\psi = \frac{\lambda}{2\pi\epsilon}\,\partial^2(\bar\psi\psi)\,,
\end{equation}
which leads to our final answer
\begin{equation}
\Delta \bar\psi\psi = \frac{\lambda}{2\pi\epsilon}\,\partial^2(\bar\psi\psi)\,,
\end{equation}
which we observe to be once again in perfect agreement with the universal expression
(\ref{universal renormalization}).

%\subsection{$[(\bar\psi\psi)^n]$}

\vspace{0.5cm}

Now let us consider ${\cal O}(\lambda)$ renormalization of the composite operator $(\bar\psi\psi)^n$,
generalizing the $n=1$ result obtained above to arbitrary integer $n\geq 1$,
\begin{align}
(\bar\psi\psi)^n = [(\bar\psi\psi)^n] + \Delta(\bar\psi\psi)^n\,,\quad
\Delta(\bar\psi\psi)^n = \frac{\lambda\,\mu^\epsilon}{2}\,(\bar\psi\psi)^n \, \int d^dx\, T_{\lambda\rho}T^{\lambda\rho}\,,
\end{align}
where in the latter expression we imply performing all possible contractions and keeping only the
singular terms in the $d\rightarrow 2$ limit. Just as in the analogous calculation of renormalization of the
$\left((\partial\phi)^2\right)^n$ in subsection \ref{subsec:partial phi n} for the scalar, it turns out that we
need to contract at most two pairs of fermions in order to obtain logarithmically divergent contributions.

%As in subsection \ref{subsec: bar psi psi} where we calculated renormalized $\Delta(\bar\psi\psi)$
As in the case of $n=1$,
it is convenient to split terms contributing to
$\Delta(\bar\psi\psi)^n$ into two groups,
\begin{equation}
\Delta(\bar\psi\psi)^n = \Delta_{(1)}(\bar\psi\psi)^n + \Delta_{(2)}(\bar\psi\psi)^n\,,
\end{equation}
where $\Delta_{(1)}(\bar\psi\psi)^n$ is composed of terms originating from contracting constituent fermions
of $(\bar\psi\psi)^n$ with one of the two stress-energy tensor factors in the vertex $T_{\lambda\rho}T^{\lambda\rho}$,
while  $\Delta_{(2)}(\bar\psi\psi)^n$ is composed of terms originating from contracting constituent fermions
of $(\bar\psi\psi)^n$ with each of the stress-energy tensor factors in that vertex.

First, consider\footnote{Here $\frac{n(n-1)}{2!}$ is the combinatorial coefficient of picking
two operators $\bar\psi\psi$ out of $n$
to make contractions, and 8 is the combined factor of symmetry due to contributing factors, such as
two possible choices of $\psi$ and $\bar\psi$
from $(\bar\psi\psi)^2$ to make a contraction, times two possible choices of $T_{\lambda\rho}$
in the vertex $T_{\lambda\rho}T^{\lambda\rho}$
to contract with, times two possible terms in $T_{\lambda\rho}$ to contract with,
which result in identical contribution due to the symmetry of $T_{\lambda\rho}$.}
\begin{align}
&\Delta_{(1)}(\bar\psi\psi)^n = n\, (\bar\psi\psi)^{n-1}\, \Delta_{(1)}(\bar\psi\psi)\\
&+\frac{\lambda}{2}\left(\frac{1}{4}\right)^2 \frac{n(n-1)}{2!} \,8\, (\bar\psi\psi)^{n-2}
\int d^dx\, \left(\bar\psi(0)\langle \psi(0)\bar\psi(x)\rangle \gamma_\lambda
\langle \partial_\rho\psi(x)\bar\psi(0)\rangle \psi(0)\, T^{\lambda\rho}
+\textrm{c.c.}\right)\,.\notag
\end{align}
As we have obtained above, $\Delta_{(1)}(\bar\psi\psi)=0$.
Keeping only singular in the $d\rightarrow 2$ limit terms, we obtain
\begin{align}
\Delta_{(1)}(\bar\psi\psi)^n &=\frac{\lambda\,n(n-1)}{32\pi^2}\,(\bar\psi\psi)^{n-2}\,
\left(\bar\psi\gamma_\mu\gamma_\lambda\gamma_\rho\psi\,\I^{\mu\alpha}
-2\bar\psi\gamma_\mu\gamma_\lambda\gamma_\nu\psi\,\I^{\mu\nu\rho\alpha}\right)\,
\partial_\alpha \, T^{\lambda\rho} +\textrm{c.c.}
\end{align}
One can verify that the first group of terms on the r.h.s. of the last expression is purely imaginary, hence
the total is zero.

Next we consider cross-contraction terms
\begin{align}
\Delta_{(2)}(\bar\psi\psi)^n =
\Delta_{(2)}^{(1)}(\bar\psi\psi)^n+
\Delta_{(2)}^{(2)}(\bar\psi\psi)^n+
\Delta_{(2)}^{(3)}(\bar\psi\psi)^n\,,
\end{align}
where we denoted
\begin{align}
\label{Delta 2 1 bar psi psi n result}
\Delta_{(2)}^{(1)}(\bar\psi\psi)^n=n\, (\bar\psi\psi)^{n-1}\, \Delta_{(2)}(\bar\psi\psi)
=\frac{\lambda n}{2\pi\epsilon}\,(\bar\psi\psi)^{n-1}\,\partial^2(\bar\psi\psi)\,,
\end{align}

\begin{align}
\label{Delta 2 2 bar psi psi intermediate}
&\Delta_{(2)}^{(2)}(\bar\psi\psi)^n=\frac{\lambda}{2}\left(\frac{1}{4}\right)^2 \frac{n(n-1)}{2!}\,4
\,(\bar\psi\psi)^{n-2}\,\int d^dx\\
&\times\left(2\left(\bar\psi\gamma^\lambda\langle \partial^\rho\psi(x)\bar\psi(0)\rangle
\psi(0)\bar\psi(0)\langle\psi(0)\bar\psi(x)\rangle\gamma_\lambda\partial_\rho\psi
+\bar\psi\gamma^\lambda\langle 
\partial^\rho\psi(x)\bar\psi(0)\rangle
\psi(0)\bar\psi(0)\langle\psi(0)\bar\psi(x)\rangle\gamma_\rho\partial_\lambda\psi\right)\right.\notag\\
&-\left.\left(
\partial^\lambda\bar\psi\gamma^\rho\langle \psi(x)\bar\psi(0)\rangle
\psi(0)\bar\psi(0)\langle \psi(0)\bar\psi(x)\rangle\gamma_\lambda\partial_\rho\psi
+\partial^\rho\bar\psi\gamma^\lambda\langle \psi(x)\bar\psi(0)\rangle
\psi(0)\bar\psi(0)\langle \psi(0)\bar\psi(x)\rangle\gamma_\lambda\partial_\rho\psi
\right)\right.\notag\\
&-\left.
\left(\bar\psi\gamma_\lambda\langle \partial_\rho\psi(x)\bar\psi(0)\rangle
\psi(0)\bar\psi(0)\langle\psi(0)\partial^\lambda\bar\psi(x)\rangle\gamma^\rho\psi
+
\bar\psi\gamma_\rho\langle \partial_\lambda\psi(x)\bar\psi(0)\rangle
\psi(0)\bar\psi(0)\langle\psi(0)\partial^\lambda\bar\psi(x)\rangle\gamma^\rho\psi
\right)\right)\notag\\
&+\textrm{c.c.}\,,\notag
\end{align}

\begin{align}
\label{Delta 2 3 bar psi psi intermediate}
&\Delta_{(2)}^{(3)}(\bar\psi\psi)^n=\frac{\lambda}{2}\left(\frac{1}{4}\right)^2 \frac{n(n-1)}{2!}\,4
\,(\bar\psi\psi)^{n-2}\,\int d^dx\\
&\times\left(\left(
\bar\psi(0)\langle \psi(0)\bar\psi(x)\rangle \gamma_\lambda\partial_\rho\psi
\bar\psi(0)\langle \psi(0)\bar\psi(x)\rangle \gamma^\lambda\partial^\rho\psi
+
\bar\psi(0)\langle \psi(0)\bar\psi(x)\rangle \gamma_\rho\partial_\lambda\psi
\bar\psi(0)\langle \psi(0)\bar\psi(x)\rangle \gamma^\lambda\partial^\rho\psi
\right)\right.\notag\\
&-\left.\left(
\bar\psi(0)\langle\psi(0)\bar\psi(x)\rangle\gamma_\lambda\partial_\rho\psi\bar\psi(0)
\langle\psi(0)\partial^\lambda\bar\psi(x)\rangle \gamma^\rho\psi
+
\bar\psi(0)\langle\psi(0)\bar\psi(x)\rangle\gamma_\lambda\partial_\rho\psi\bar\psi(0)
\langle\psi(0)\partial^\rho\bar\psi(x)\rangle \gamma^\lambda\psi
\right)\right.\notag\\
&+\left.\left(
\bar\psi\gamma_\lambda\langle\partial_\rho\psi(x)\bar\psi(0)\rangle\psi(0)
\bar\psi\gamma^\lambda\langle\partial^\rho\psi(x)\bar\psi(0)\rangle\psi(0)
+
\bar\psi\gamma_\rho\langle\partial_\lambda\psi(x)\bar\psi(0)\rangle\psi(0)
\bar\psi\gamma^\lambda\langle\partial^\rho\psi(x)\bar\psi(0)\rangle\psi(0)
\right)\right.\notag\\
&-\left.\left(
\bar\psi\gamma_\lambda\langle\partial_\rho\psi(x)\bar\psi(0)\rangle\psi(0)
\partial^\lambda\bar\psi(x)\gamma^\rho\langle\psi(x)\bar\psi(0)\rangle\psi(0)
+
\bar\psi\gamma_\lambda\langle\partial_\rho\psi(x)\bar\psi(0)\rangle\psi(0)
\partial^\rho\bar\psi(x)\gamma^\lambda\langle\psi(x)\bar\psi(0)\rangle\psi(0)
\right)\right)\notag\\
&+\textrm{c.c.}\notag
\end{align}
Extracting the singular in the $d\rightarrow 2$ limit terms and simplifying
these ${\cal O}(\lambda)$ expressions using ${\cal O}(1)$ equations of motion, we obtain
(we relegate details to appendix \ref{appendix: details of bar psi psi to n})
\begin{align}
\label{Delta 2 2 bar psi psi n result}
\Delta_{(2)}^{(2)}(\bar\psi\psi)^n &= \frac{\lambda n(n-1)}{\pi\epsilon}\,
(\bar\psi\psi)^{n-2}\, \partial_\mu\bar\psi\psi\,\bar\psi\partial^\mu\psi\,,\\
\label{Delta 2 3 bar psi psi n result}
\Delta_{(2)}^{(3)}(\bar\psi\psi)^n &= \frac{\lambda n(n-1)}{2\pi\epsilon}\,
(\bar\psi\psi)^{n-2}\, \left(\bar\psi\partial_\mu\psi\,\bar\psi\partial^\mu\psi
+\partial_\mu\bar\psi\psi\,\partial^\mu\bar\psi\psi\right)\,.
\end{align}
Combining (\ref{Delta 2 1 bar psi psi n result}), (\ref{Delta 2 2 bar psi psi n result}), (\ref{Delta 2 3 bar psi psi n result})
and the fact that $\Delta_{(1)} (\bar\psi\psi)^n=0$
we obtain
\begin{equation}
\Delta(\bar\psi\psi)^n=\frac{\lambda}{2\pi\epsilon}\partial^2(\bar\psi\psi)^n\,,
\end{equation}
in agreement with the universal expression
(\ref{universal renormalization}).

\subsection{$[\bar\psi\gamma_\mu\psi]$}

We now briefly discuss renormalization of the conserved $U(1)$ current $\bar\psi\gamma_\mu\psi$.
As reviewed in Introduction section, Noether currents are expected not to be renormalized due to Ward identity.
In case of stress-energy tensor a possible exception to this statement is given by identically
conserved improvement counterterms (which we, however, derived to be vanishing for the considered models).
In case of a free fermion deformed by $T\bar T$ a potential counterpart exception can be written down as
\begin{equation}
\frac{1}{\epsilon}\,\lambda\,\left(\partial_\mu\partial_\nu-\delta_{\mu\nu}\,\partial^2\right)\,(\bar\psi\gamma^\nu\psi)
\sim\frac{1}{\epsilon}\,\lambda\,\partial^2\,(\bar\psi\gamma_\mu\psi)\,.
\end{equation}
Yet we 
have explicitly verified $\Delta(\bar\psi\gamma_\mu\psi)=0$ at one-loop level.\footnote{We established this without the use of equations of motion.}

To avoid burdening the reader with excessive and
repetitive details of calculation, we skip detailed explanations, and refrain to pointing out that at every single
step the calculation is analogous to calculation of $[\bar\psi\psi]$ in subsection \ref{subsec: bar psi psi}.

We then point out that the statement $\Delta(\bar\psi\gamma_\mu\psi)=0$ still fits
the universal expression (\ref{universal renormalization}). Indeed, in two dimensions
one obtains $\lambda\,\partial^2\,(\bar\psi\gamma_\mu\psi) = 0$ once the ${\cal O}(\lambda^0)$
e.o.m. for $\psi$ is taken into account.

\subsection{$[\bar\psi\gamma_\mu\partial_\nu\psi]$}

Consider renormalization of the composite tensor operator $\bar\psi\gamma_\mu\partial_\nu\psi$,
\begin{align}
\bar\psi\gamma_\mu\partial_\nu\psi = [\bar\psi\gamma_\mu\partial_\nu\psi] 
+\Delta \bar\psi\gamma_\mu\partial_\nu\psi\,,
\quad \Delta \bar\psi\gamma_\mu\partial_\nu\psi=
 \frac{\lambda\,\mu^\epsilon}{2}\,\bar\psi\gamma_\mu\partial_\nu\psi \int d^dx \, T_{\lambda\rho} T^{\lambda\rho}\,,
\end{align}
where in $\Delta \bar\psi\gamma_\mu\partial_\nu\psi$ we take all possible contractions and retain singular 
in $d\rightarrow 2$ limit terms only.
As before, when calculating renormalization of composite operators, we choose to position
the composite operator itself at the origin, $x=0$, without the loss of generality. 
To calculate $\Delta \bar\psi\gamma_\mu\partial_\nu\psi$ we find
it convenient to split possible contributions to it into two groups,
\begin{equation}
\Delta \bar\psi\gamma_\mu\partial_\nu\psi = \Delta _{(1)}\bar\psi\gamma_\mu\partial_\nu\psi
+\Delta _{(2)}\bar\psi\gamma_\mu\partial_\nu\psi\,.
\end{equation}

First, we consider singular terms originating from contracting the constituent fermion fields of the composite
operator $\bar\psi\gamma_\mu\partial_\nu\psi$ with any one of the two factors $T_{\lambda\rho}$
in the interaction vertex $T_{\lambda\rho}T^{\lambda\rho}$. Taking into account symmetric properties
of the stress-energy tensor we obtain
\begin{equation}
\label{Delta bar psi gamma mu d nu psi original}
\Delta _{(1)} \bar\psi\gamma_\mu\partial_\nu\psi = - \frac{1}{2}\lambda\mu^\epsilon
\int d^dx\,\textrm{Tr}\,\left(\langle \partial_\nu\psi(0)\bar\psi(x)\rangle \gamma_\lambda \langle
\partial_\rho\psi(x)\bar\psi(0)\rangle\gamma_\mu+\textrm{c.c.}\right)\,T^{\lambda\rho}(x)\,.
\end{equation}
Using here \footnote{As discussed above,
we set $d=2$ in the factors which are not going to contribute to simple poles, which are
expected to occur in the linear order of the perturbation theory.}
\begin{equation}
\langle \partial_\nu\psi(x) \bar\psi(0)\rangle =  \frac{1}{2\pi}\,\frac{1}{|x|^d}\,
\left(\frac{2x^\mu x_\nu \gamma_\mu}{|x|^2} - \gamma_\nu\right)\,,
\end{equation}
we notice that on the r.h.s. of (\ref{Delta bar psi gamma mu d nu psi original}) we have terms of the form
\begin{equation}
\textrm{Tr} (\gamma_\mu\gamma_\nu\gamma_\lambda\gamma_\rho) =2(\delta_{\mu\nu}
\delta_{\lambda\rho} - \delta_{\rho\nu}\delta_{\mu\lambda} + \delta_{\lambda\nu}\delta_{\mu\rho})\,,
\end{equation}
which are real-valued, and therefore (\ref{Delta bar psi gamma mu d nu psi original}) can be re-written as
\begin{equation}
\label{Delta bar psi gamma mu d nu psi original simple}
\Delta _{(1)} \bar\psi\gamma_\mu\partial_\nu\psi = - \lambda
\int d^dx\,\textrm{Tr}\,\left(\langle \partial_\nu\psi(0)\bar\psi(x)\rangle \gamma_\lambda \langle
\partial_\rho\psi(x)\bar\psi(0)\rangle\gamma_\mu\right)\,T^{\lambda\rho}(x)\,.
\end{equation}

Expanding the integrated $T^{\lambda\rho}(x)$ operator around $x=0$,
and keeping only logarithmically divergent at short distances terms, while taking
into account that the stress-energy tensor is symmetric and traceless, we obtain
\begin{equation}
\Delta _{(1)} \bar\psi\gamma_\mu\partial_\nu\psi = \frac{\lambda}{2\pi^2} \,\left(
\I^{\alpha\beta\rho}_{\quad\;\;\;\mu}\,\partial_\alpha\partial_\beta \, T_{\nu\rho}
+ \I^{\alpha\beta\rho}_{\quad\;\;\;\nu}\,\partial_\alpha\partial_\beta \, T_{\mu\rho}
+\delta_{\mu\nu}\,\I^{\alpha\beta\lambda\rho}\,\partial_\alpha\partial_\beta\, T_{\lambda\rho}
-4\,\I_{\mu\nu}^{\quad \alpha\beta\lambda\rho}\,\partial_\alpha\partial_\beta\, T_{\lambda\rho}\right)\,,
\end{equation}
where we have identified the master integrals (\ref{master integral general definition}).
Using (\ref{I4 expression}), (\ref{I6 expression}) one can demonstrate that 
\begin{equation}
\label{Delta 1 bar psi gamma mu d nu psi result}
\Delta _{(1)} \bar\psi\gamma_\mu\partial_\nu\psi = 0.
\end{equation}

The second contribution to $\Delta \bar\psi\gamma_\mu\partial_\nu\psi$ originates
from all possible cross-contractions between the constituent fermions of the composite
operator $\bar\psi\gamma_\mu\partial_\nu\psi(0)$ with each of the two factors $T_{\lambda\rho}$
in the interaction vertex $T_{\lambda\rho}T^{\lambda\rho}$,
\begin{align}
\label{Delta 2 bar psi gamma mu d nu psi eight total}
&\Delta_{(2)} \bar\psi\gamma_\mu\partial_\nu\psi=
\frac{\lambda}{2}\left(\frac{1}{4}\right)^2\cdot 4\\
&\times\left(\int d^dx\,\left(\bar\psi \gamma^\lambda \langle \partial^\rho \psi(x)\bar\psi (0)\rangle
\gamma_\mu \langle \partial_\nu\psi(0)\bar\psi(x)\rangle \gamma_\lambda\partial_\rho\psi
+
\bar\psi \gamma^\lambda \langle \partial^\rho \psi(x)\bar\psi (0)\rangle
\gamma_\mu \langle \partial_\nu\psi(0)\bar\psi(x)\rangle \gamma_\rho\partial_\lambda\psi
\right)\right.\notag\\
&-\left.
\int d^dx\,\left(\bar\psi \gamma_\lambda \langle \partial_\rho \psi(x)\bar\psi (0)\rangle
\gamma_\mu \langle \partial_\nu\psi(0)\partial^\lambda\bar\psi(x)\rangle \gamma^\rho\psi
+\partial^\lambda\bar\psi \gamma^\rho \langle  \psi(x)\bar\psi (0)\rangle
\gamma_\mu \langle \partial_\nu\psi(0)\bar\psi(x)\rangle \gamma_\lambda\partial_\rho\psi
\right)\right.\notag\\
&-\left.
\int d^dx\,\left(\bar\psi \gamma_\lambda \langle \partial_\rho \psi(x)\bar\psi (0)\rangle
\gamma_\mu \langle \partial_\nu\psi(0)\partial^\rho\bar\psi(x)\rangle \gamma^\lambda\psi
+\partial^\rho\bar\psi \gamma^\lambda \langle  \psi(x)\bar\psi (0)\rangle
\gamma_\mu \langle \partial_\nu\psi(0)\bar\psi(x)\rangle \gamma_\lambda\partial_\rho\psi
\right)\right.\notag\\
&+\left.
\int d^dx\left(
\partial^\rho\bar\psi\gamma^\lambda\langle \psi(x)\bar\psi(0)\gamma_\mu
\langle \partial_\nu\psi(0)\partial_\rho\bar\psi(x)\rangle \gamma_\lambda\psi
+\partial^\lambda\bar\psi\gamma^\rho\langle \psi(x)\bar\psi(0)\rangle\gamma_\mu
\langle \partial_\nu\psi(0)\partial_\rho\bar\psi(x)\rangle \gamma_\lambda\psi
\right)\right)\,.\notag
\end{align}
While this is somewhat laborious, one can show that each of the eight terms contributing to
$\Delta_{(2)} \bar\psi\gamma_\mu\partial_\nu\psi$ is individually finite in $d\rightarrow 2$ limit.
As an example, consider the first term
\begin{align}
\Delta_{(2)}^{(1)} \bar\psi\gamma_\mu\partial_\nu\psi&\equiv
\frac{\lambda}{8}\,\int d^dx\,\bar\psi \gamma^\lambda \langle \partial^\rho \psi(x)\bar\psi (0)\rangle
\gamma_\mu \langle \partial_\nu\psi(0)\bar\psi(x)\rangle \gamma_\lambda\partial_\rho\psi\\
&=\frac{\lambda}{64\pi^2}\left(\I^{\alpha\beta}\,\partial_\alpha\partial_\beta\,
(\bar\psi\gamma^\lambda\gamma^\rho\gamma_\mu\gamma_\nu\gamma_\lambda\partial_\rho\psi)
-2\,\I^{\alpha\beta\rho\sigma}\,\partial_\alpha\partial_\beta\,(\bar\psi
\gamma^\lambda\gamma_\sigma\gamma_\mu\gamma_\nu\gamma_\lambda\partial_\rho\psi)\right.\notag\\
&-\left.2\,\I^{\alpha\beta\sigma}_{\quad\;\;\;\nu}\,\partial_\alpha\partial_\beta\, (\bar\psi
\gamma^\lambda\gamma^\rho\gamma_\mu\gamma_\sigma\gamma_\lambda\partial_\rho\psi)
+4\,\I^{\alpha\beta\rho\sigma\tau}_{\quad\quad\;\;\nu}\,\partial_\alpha\partial_\beta\,
(\bar\psi\gamma^\lambda\gamma_\sigma\gamma_\mu\gamma_\tau\gamma_\lambda\partial_\rho\psi)\right)\,.
\notag
\end{align}
This expression can be quickly simplified in \textit{Mathematica} using (\ref{I2 expression}),
(\ref{I4 expression}), (\ref{I6 expression}), rendering $\Delta_{(2)}^{(1)} \bar\psi\gamma_\mu\partial_\nu\psi=0$.
Analogous calculation will show that the other contributions to (\ref{Delta 2 bar psi gamma mu d nu psi eight total})
vanish as well. Combining this with (\ref{Delta 1 bar psi gamma mu d nu psi result}) we conclude that
\begin{equation}
\label{Delta bar psi gamma mu d nu psi result}
\Delta \bar\psi\gamma_\mu\partial_\nu\psi = 0\,.
\end{equation}

The result (\ref{Delta bar psi gamma mu d nu psi result}) for the linear order correction to the
renormalized $[\bar\psi\gamma_\mu\partial_\nu\psi]$ actually agrees with our universal
expression (\ref{universal renormalization}). This is due to the fact that once the ${\cal O}(\lambda^0)$
equations of motion $\gamma^\mu\partial_\mu\psi = 0$ are imposed, the ${\cal O}(\lambda)$
contribution on the r.h.s. of (\ref{universal renormalization}) can be simplified in two dimensions,
\begin{equation}
\partial^2 (\bar\psi\gamma_\mu\partial_\nu\psi) = 0\,.
\end{equation}

\subsection{$[T_{\mu\nu}]$}

Using expression for the stress-energy tensor (\ref{fermion stress energy tensor})
and eq. (\ref{Delta bar psi gamma mu d nu psi result}) derived in previous subsection
 for the renormalization of $[\bar\psi\gamma_\mu\partial_\nu\psi]$
we conclude that
\begin{align}
[T_{\mu\nu}] &= T_{\mu\nu} +{\cal O}(\lambda^2)\,,\\
\partial^2 \, T_{\mu\nu} &= 0\,,
\end{align}
which agree with the universal relation (\ref{universal renormalization}).
This also agrees with the general argument of \cite{Cardy:2019qao}, which
can be applied here to argue that $[T_{\mu\nu}]=T_{\mu\nu}$ to all orders in $\lambda$.

\section{Massive case}
\label{sec: massive case}

In previous sections, we extensively explored the renormalization of various composite operators
in free massless scalar and fermionic theories deformed by the $T\bar T$ operator.
By carrying our explicit perturbative calculation to linear order
in the $T\bar T$ coupling, we have demonstrated that all of the considered
operators are renormalized by a universal counter-term (\ref{universal renormalization}).
It has been suggested in \cite{Cardy:2019qao} that such a universal behavior is a general
feature in the $T\bar T$-deformed CFTs.

It is therefore crucial to explore the fate of the composite operators
in the non-conformal field theories deformed by the $T\bar T$ operator. 
Indeed, dimensional considerations suggest that QFTs with a mass scale
could exhibit a richer structure of counterterms compared to (\ref{universal renormalization}).
In this section, we intend to investigate this question by considering $T\bar T$
deformation of a free massive scalar theory.

One of the important results which we have obtained for the $T\bar T$-deformed
free massless scalar and fermionic theories in sections,~\ref{sec:scalar}~\ref{sec:fermion}
was that the improvement (`gravitational counterterm') contribution, typically appearing in the
renormalization of the stress-energy tensor, is trivial in those theories.
In fact, we have demonstrated that the stress-energy tensor does not receive any counterterms
at the linear order in the $T\bar T$ coupling. We expect this conclusion will not persist in the case
of $T\bar T$ deformed massive theories. In fact, it is well known that even a free massive
scalar field requires a cosmological constant counterterm, which contributes the identity operator to the renormalizaton
of the stress-energy tensor.

\iffalse
This section is complementary to our main results concerning CFTs,
and can be skipped without affecting the reader's understanding of the rest of this paper.
Here we intend to venture outside of the scope of the universal composite operator
renormalization in conformal field theories, which we have been investigating so far.
Recall that one of our results above states that the stress-energy tensor, viewed as a composite operator
in the free massless scalar and fermion theories deformed by the
$T\bar T$ operator, is not renormalized to the linear order of the $T\bar T$ coupling $\lambda$.
Specifically, we have demonstrated that
the improvement (`gravitational counterterm') contribution, typically appearing in the
renormalization of the stress-energy tensor, is trivial in the considered theories.
This section is motivated by the question of the extent of this fate of the improvement
counterterms outside of the
scope of conformal fields theories.
\fi

Recall that due to the Ward identity the possible renormalization of the stress-energy tensor is
either a total derivative, \textit{i.e.}, an improvement term (which formally can be seen as arising
from the gravitational counterterm), or identity operator, or it is proportional to the stress-energy tensor itself.
In this section we will consider the free massive scalar deformed by the $T\bar T$
operator, and study it to the linear order in the $T\bar T$ coupling $\lambda$,
\footnote{This action is written in terms of finite physical mass $m$
and field $\phi$, which are related to the bare $m_0$, $\phi_0$ via the standard equations
\begin{equation}
\phi_0=\phi\,\sqrt{1+\delta_\phi}\,,\qquad m_0^2(1+\delta_\phi) = m^2(1+\delta_m)\,,
\end{equation}
where the counterterms $\delta_{m,\phi}$ are non-trivial in the massive case even at the linear order in coupling
$\lambda$.}
\begin{align}
I&=\int d^dx\, \left(\frac{1}{2}\, \partial_\mu \phi \partial ^\mu\phi
+\frac{1}{2} \, m^2\, \phi^2 + \frac{1}{2}\,\delta_\phi \, \partial _\mu\phi \partial^\mu \phi 
+\frac{1}{2}\,m^2\,\delta_m \, \phi^2\right)\notag\\
&+\frac{\lambda \mu^{\epsilon}}{4}\,\int d^dx\, \left(m^4\phi^4
-((\partial  \phi)^2)^2\right)\,,
\label{renormalized massive action}
\end{align}
where the $T\bar T$ operator at ${\cal O}(\lambda^0)$ is given by
\begin{equation}
\label{massive TT0 in terms of phi}
T\bar T_0 = \frac{1}{4}\left(m^4\phi^4
-((\partial  \phi)^2)^2\right)\,,
\end{equation}
and $d=2-\epsilon$. We argue that the ${\cal O}(\lambda)$ renormalization of the stress-energy
tensor is given by
\begin{equation}
\label{massive scalar renormalization of stress-energy tensor}
[T_{\mu\nu}] = \left( 1 - \frac{3\lambda m^2}{2\pi\epsilon}  \right)
\left(T_{\mu\nu}+\frac{m^2}{4\pi\epsilon}\,\delta_{\mu\nu} \, \mathbb{I}\right) + {\cal O}(\lambda^2)\,.
\end{equation}
Notice that the form (\ref{massive scalar renormalization of stress-energy tensor})
is consistent with the Ward identity, and it does not include the improvement counterterm contribution.
However, the $T\bar T$ coupling $\lambda$ induces a multiplicative renormalization of the
stress-energy tensor.

We now proceed to the derivation of (\ref{massive scalar renormalization of stress-energy tensor}).
First of all, the mass and wave-function renormalization counterterm contributions at ${\cal O}(\lambda)$ order
in the massive action (\ref{renormalized massive action}) are defined by
\begin{align}
\label{massive scalar counterterms}
\delta _m = -\frac{3\lambda m^2}{2\pi \epsilon}\,,\qquad
\delta _ \phi = - \frac{\lambda m^2}{\pi \epsilon}\,,
\end{align}
as, \textit{e.g.}, can be readily seen by renormalizing the two-point function
\begin{align}
\label{propagator with loop corrections}
\langle \phi(p)\phi(q)\rangle &=(2\pi)^d\delta (p+q)\,
\left(\frac{1}{p^2+m^2} + \frac{-m^2\delta _m -p^2 \delta _\phi}{(p^2+m^2)^2}\right)\\
&-\frac{\lambda \mu^{\epsilon}}{4}\,\int d^dx\, \left(m^4\,\langle \phi^4(x)\phi(p)\phi(q)\rangle
-\langle((\partial  \phi(x))^2)^2\phi(p)\phi(q)\rangle\right)\,,
\end{align}
and using
\begin{align}
\label{phi squared free vev}
\langle \phi^2\rangle = \frac{(m^2)^{-\frac{\epsilon}{2}}}{2\pi\epsilon}\,,\quad
\langle \partial_\mu\phi\partial_\nu\phi\rangle = -\frac{m^2}{4\pi\epsilon} \, \delta_{\mu\nu} \,,\quad
\langle (\partial \phi)^2\rangle = - \frac{m^2}{2\pi\epsilon}\,.
\end{align}

The canonical stress-energy tensor to the linear order ${\cal O}(\lambda)$, following from
the renormalized action (\ref{renormalized massive action}), is given by
\begin{align}
T_{\mu\nu}  = \partial_\mu\phi_0\partial_\nu\phi_0 (1-\lambda(\partial\phi_0)^2)-
\frac{1}{2}\,\delta_{\mu\nu}\,\left((\partial\phi_0)^2+m_0^2\phi_0^2+2\lambda T\bar T_0\right)\,,
\label{stress energy tensor for massive scalar}
\end{align}
where we re-absorbed the counterterms into the bare mass $m_0$ and field $\phi_0$. 
We will be using the stress-energy tensor (\ref{stress energy tensor for massive scalar})
expressed in terms of bare $m_0$, $\phi_0$, but to lighten the notation we will skip putting the zero subscript
whenever the resulting expression does not get affected to the linear order in $\lambda$.

Using (\ref{phi squared free vev}), (\ref{stress energy tensor for massive scalar}) we immediately
notice that to ${\cal O}(\lambda^0)$, \textit{i.e.}, in the free theory, due to non-vanishing mass, the renormalized 
stress-energy tensor is determined by the cosmological constant renormalization,
\begin{equation}
\label{free massive scalar renormalization of stress-energy tensor}
[T_{\mu\nu}] =T_{\mu\nu}+\frac{m^2}{4\pi\epsilon}\,\delta_{\mu\nu} \, \mathbb{I}+ {\cal O}(\lambda)\,,
\end{equation}
in agreement with (\ref{massive scalar renormalization of stress-energy tensor}).
The rest of this section is dedicated to derivation of the ${\cal O}(\lambda)$
contribution to (\ref{massive scalar renormalization of stress-energy tensor}).
Specifically, we will extend the derivation in section \ref{sec:scalar} and find the terms contributing
to $[T_{\mu\nu}]$ due to non-vanishing mass $m$. We will denote the corresponding ${\cal O}(\lambda)$ contributions
to renormalized stress-energy tensor $[T_{\mu\nu}]$ with $\Delta_{(m)}(T_{\mu\nu})$, that is,
$T_{\mu\nu} = [T_{\mu\nu}] + \Delta_{(m)}(T_{\mu\nu})+{\cal O}(\lambda^2)+\dots$,
where ellipsis stand for contributions at $m=0$. Using (\ref{stress energy tensor for massive scalar})
we obtain to ${\cal O}(\lambda)$
\begin{align}
\label{delta m T total}
\Delta_{(m)}(T_{\mu\nu}) &=
\Delta_{(m)}(\partial_\lambda\phi\partial_\rho\phi)\,
\left(\delta_\mu^\lambda\delta_\nu^\rho - \frac{1}{2}\,\delta^{\lambda\rho}\,\delta_{\mu\nu}\right)
-\lambda\,\Delta_{(m)}(\partial_\mu\phi\partial_\nu\phi (\partial\phi)^2)\\
&-\frac{1}{2}\,m^2\,\delta_{\mu\nu}\,\Delta(\phi^2)
-\lambda\,\delta_{\mu\nu}\,\Delta_{(m)}(T\bar T_0)\,,
\end{align}
where renormalization $\Delta_{(m)}$ of various operators on the r.h.s. is to be taken up to the
appropriate order in $\lambda$, so that $\Delta_{(m)}(T_{\mu\nu})$ is of the linear order in $\lambda$.
Notice that $\Delta(\phi^2)$ in (\ref{delta m T total}) includes both the zero-mass and
(possible) finite-mass contributions.

From the action (\ref{renormalized massive action}), (\ref{massive scalar counterterms})
we can derive
the renormalized composite operator $[T\bar T_0]$ to ${\cal O}(\lambda)$,
\begin{equation}
\label{composite TT renormalization}
[T\bar T_0] = \mu^{-\epsilon}\,\frac{\partial L}{\partial \lambda}
=T\bar T_0 - \frac{m^2}{4\pi\epsilon} \, \left(2(\partial\phi)^2 +3
m^2\phi^2\right)\;\;\Rightarrow\;\;\Delta_{(m)}(T\bar T_0) = \frac{m^2}{4\pi\epsilon} \, \left(2(\partial\phi)^2 +3
m^2\phi^2\right) \,.
\end{equation}
As a consistency check,
the relation (\ref{composite TT renormalization})
can alternatively be obtained by directly renormalizing (\ref{massive TT0 in terms of phi}) at ${\cal O}(\lambda^0)$
order using (\ref{phi squared free vev}) and (here we skip for now renormalization terms proportional to
 the identity operator $\mathbb{I}$, these will be restored in the final expression)
\begin{align}
\phi^4 &=[\phi^4] +6 \phi^2\,\langle \phi^2\rangle +\dots
=[\phi^4] +\frac{3}{\pi\epsilon}\,\phi^2+\dots\,,\\
((\partial \phi) ^2)^2 &=
[((\partial \phi) ^2)^2]
+2\langle (\partial\phi)^2\rangle \,(\partial\phi)^2
+4\langle \partial_\alpha\phi\partial_\beta\phi\rangle \partial^\alpha\phi\partial^\beta\phi+\dots
=[((\partial \phi) ^2)^2]-\frac{2m^2}{\pi\epsilon}(\partial\phi)^2+\dots\,.\notag
\end{align}

Similarly, to ${\cal O}(\lambda^0)$ we derive 
\begin{align}
\Delta_{(m)}(\partial_\mu\phi\partial_\nu\phi (\partial\phi)^2) &=
2\langle \partial_\mu \phi\partial_\alpha\phi\rangle \partial_\nu\phi\partial^\alpha\phi
+2\langle \partial_\nu \phi\partial_\alpha\phi\rangle \partial_\mu\phi\partial^\alpha\phi
+\langle \partial_\mu\phi\partial_\nu\phi\rangle (\partial\phi)^2
+ \partial_\mu\phi\partial_\nu\phi\langle (\partial\phi)^2\rangle\notag\\
&=-\frac{3m^2}{2\pi\epsilon}\,\partial_\mu\phi\partial_\nu\phi
-\frac{m^2}{4\pi\epsilon}\,\delta_{\mu\nu}\,(\partial\phi)^2\,.
\label{dphidphi dphi squared renormalization}
\end{align}

The mass contribution to
renormalization of $\phi^2$ at ${\cal O}(\lambda)$, which needs to be added to the expression
(\ref{Delta phi ^ n preliminary}),
evaluated for $n=2$, is given by
\begin{equation}
\label{Delta m for phi^2}
\Delta_{(m)}(\phi^2)=-\frac{\lambda \mu ^\epsilon}{4} \,m^4\,\phi^2\int d^dx\,
\phi^4 = -3\lambda \,m^4\,\int d^dx\,\langle \phi(0)\phi(x)\rangle^2\,\phi^2 \,,
\end{equation}
where we recall that for the free massive scalar the propagator is given by
\begin{equation}
\langle \phi(0)\phi(x)\rangle=\frac{1}{(2\pi)^\frac{d}{2}}\,\left(\frac{|m|}{|x|}\right)^\frac{d-2}{2}\,
K_\frac{d-2}{2}\left(|m||x|\right)\,.
\end{equation}
It can be seen that (\ref{Delta m for phi^2}) does not contain any logarithmic divergencies
(is regular in $d\rightarrow 2$ limit),
\begin{equation}
\label{delta m phi squared}
\Delta_{(m)}(\phi^2) = 0\,.
\end{equation}

Next, extending the calculation of subsection \ref{subsec: partial mu phi partial nu phi},
we consider mass contributions to renormalization
of $\partial_\mu\phi\partial_\nu\phi$,
\begin{align}
\label{Delta m for d mu phi d nu}
&\Delta_{(m)}(\partial_\mu\phi\partial_\nu\phi)=
 \frac{\lambda}{4}\,
\int d^dx \,\left( 4\, \partial_\mu\partial_\alpha \langle \phi(0)\phi(x)\rangle
\partial_\nu\partial^\alpha \langle \phi(0)\phi(x)\rangle (\partial\phi(x))^2\right.\\
&+\left. 8\, \partial_\mu\partial_\alpha \langle \phi(0)\phi(x)\rangle
\partial_\nu\partial_\beta \langle \phi(0)\phi(x)\rangle \partial^\alpha\phi(x)\partial^\beta\phi(x) \right)
\Bigg|_{m}
-3m^4\,\int d^dx\,\partial_\mu\langle \phi(0)\phi(x)\rangle \partial_\nu\langle \phi(0)\phi(x)\rangle \phi^2\,.\notag
\end{align}
Here the first two terms on the r.h.s. are recognized as having already appeared
in the analogous calculation (\ref{d mu phi d nu phi start}) in the massless case.
Therefore now in these terms we only retain contributions due to the mass $m$, as we indicated
with the subscript $m$. The last term in (\ref{Delta m for d mu phi d nu}) is due to the $m^4$ term
in the $T\bar T$ operator (\ref{massive TT0 in terms of phi}). We denote these two groups of contributions 
correspondingly as $\Delta_{(m)}^{(1,2)}(\partial_\mu\phi\partial_\nu\phi)$,
\begin{equation}
\Delta_{(m)}(\partial_\mu\phi\partial_\nu\phi) = \Delta_{(m)}^{(1)}(\partial_\mu\phi\partial_\nu\phi)
+\Delta_{(m)}^{(2)}(\partial_\mu\phi\partial_\nu\phi)\,.
\end{equation}

Using the expansion around zero mass,
\begin{align}
\partial_\mu\partial_\alpha \langle \phi(0)\phi(x)\rangle &= \frac{\Gamma\left(\frac{d+2}{2}\right)}{\pi^\frac{d}{2}}
\frac{1}{|x|^{d+2}}\,\left(x_\alpha x_\mu -\frac{\delta_{\alpha\mu}}{d}\,|x|^2\right)
\left(1-\frac{m^2|x|^2}{2d}\right)\label{ddD mass 1}\\
&+\frac{\delta_{\mu\alpha}}{4d \pi^\frac{d}{2}}\frac{m^2 |x|^2}{|x|^d}
\left(\Gamma\left(\frac{d-2}{2}\right)
+\Gamma\left(\frac{2-d}{2}\right)\frac{(m|x|)^{d-2}}{2^{d-2}}\right)\,,\label{ddD mass 2}
\end{align}
we obtain
\begin{equation}
\Delta_{(m)}^{(1)}(\partial_\mu\phi\partial_\nu\phi) = 
\Delta_{(m)}^{(1,1)}(\partial_\mu\phi\partial_\nu\phi) + \Delta_{(m)}^{(1,2)}(\partial_\mu\phi\partial_\nu\phi)\,,
\end{equation}
where $\Delta_{(m)}^{(1,1)}(\partial_\mu\phi\partial_\nu\phi)$ is due to the first line in (\ref{ddD mass 2})
and $\Delta_{(m)}^{(1,2)}(\partial_\mu\phi\partial_\nu\phi)$ is due to the second line in (\ref{ddD mass 2}).
Specifically,
\begin{align}
\notag
\Delta_{(m)}^{(1,1)}(\partial_\mu\phi\partial_\nu\phi)
 &=-\frac{\lambda m^2}{2\pi^2}\,((\partial\phi)^2\delta^{\alpha\beta} + 2\partial^\alpha\phi\partial^\beta\phi)
 \left(\I_{\mu\nu\alpha\beta} - \frac{1}{2}\,\I_{\mu\alpha}\delta_{\nu\beta}
 - \frac{1}{2}\,\I_{\nu\alpha}\delta_{\mu\beta}+\frac{1}{4}\delta_{\mu\alpha}\delta_{\nu\beta}\right)\\
 &= -\frac{\lambda m^2}{2\pi\epsilon}(\partial\phi)^2\delta_{\mu\nu}\,,
\end{align}
and
\begin{align}
&\Delta_{(m)}^{(1,2)}(\partial_\mu\phi\partial_\nu\phi)=\frac{\lambda m^2}{4\pi^2}\,((\partial\phi)^2\delta^\alpha_\nu+2\partial^\alpha\phi\partial_\nu\phi)\\
&\times\left(\left(I_{\mu\alpha}-\frac{\delta_{\mu\alpha}}{d}I\right)
\Gamma\left(\frac{d-2}{2}\right) + \frac{m^{d-2}}{2^{d-2}}\Gamma\left(\frac{2-d}{2}\right)
\int d^dx\frac{1}{|x|^{d+2}}\left(x_\mu x_\alpha -\frac{\delta_{\mu\alpha}}{d}|x|^2\right)\right)\notag\\
&=\frac{\lambda m^2}{4\pi^2}\,((\partial\phi)^2\delta^\alpha_\nu+2\partial^\alpha\phi\partial_\nu\phi)
 \int\frac{d^dx}{|x|^{2d}}\,\left(x_\mu x_\alpha -\frac{\delta_{\mu\alpha}}{d}|x|^2\right)
\left(\Gamma\left(\frac{d-2}{2}\right) + \frac{(m|x|)^{d-2}}{2^{d-2}}\Gamma\left(\frac{2-d}{2}\right)\right)\,.\notag
\end{align}
Here gamma-functions cancel each other's singularities at $d=2$, and as a result $\Delta_{(m)}^{(1,2)}$
does not have a singularity at $d=2$.
Finally, for the calculation of $\Delta_{(m)}^{(2)}(\partial_\mu\phi\partial_\nu\phi)$,
corresponding to the last term on the r.h.s. of (\ref{Delta m for d mu phi d nu}),
it is sufficient to use the massless propagator (\ref{first derivative of phi massless propagator}),
which gives
\begin{equation}
\Delta_{(m)}^{(2)} (\partial_\mu\phi\partial_\nu\phi)= -\frac{3\lambda m^4}{4\pi^2}\,\I_{\mu\nu}\,\phi^2
= -\frac{3\lambda m^4}{4\pi\epsilon}\,\delta_{\mu\nu}\,\phi^2\,.
\end{equation}

Combining everything together we obtain
\begin{equation}
\label{delta m d mu phi d nu phi total}
\Delta_{(m)}(\partial_\mu\phi\partial_\nu\phi) = - \frac{\lambda m^2}{4\pi\epsilon}\,
(3m^2\phi^2+2(\partial\phi)^2)\,\delta_{\mu\nu}\,.
\end{equation}
Using (\ref{Delta phi to n before eom}) for $n=2$,
(\ref{composite TT renormalization}), (\ref{dphidphi dphi squared renormalization}),
(\ref{delta m phi squared}), (\ref{delta m d mu phi d nu phi total}) in (\ref{delta m T total})
we obtain
\begin{equation}
\Delta_{(m)}(T_{\mu\nu}) = \frac{3\lambda m^2}{2\pi\epsilon}\, T_{\mu\nu}+{\cal O}(\lambda^2)\,,
\end{equation}
which completes our derivation of (\ref{massive scalar renormalization of stress-energy tensor}).

\section{Discussion}

In this paper we considered perturbative renormalization of the composite operators in the $T\bar T$-deformed two-dimensional free field theories. In the massless case renormalization of various operators satisfies (\ref{universal renormalization}). This universal relation holds true regardless of whether the operator is spinless or not, and whether it is a primary or a descendant. Our explicit calculations agree with \cite{Cardy:2019qao}, where a generic $T\bar T$ deformed CFT was studied. However, in the massive case there is no universal formula to compare to. Hence, it would be interesting to derive such a formula for a generic $T\bar T$-deformed gapped quantum field theory.  

Conserved Noether currents correspond to a particularly interesting class of composite operators
whose renormalization one can study. Due to the Ward identities an allowed divergent structures of the Noether current -- the so-called improvement counterterms \cite{Callan:1970ze} --  are separately conserved. The stress-energy tensor is an example of the Noether current where improvement terms are directly related to the gravitational counterterms induced in a theory coupled to a curved background  \cite{Callan:1970ze,Collins:1965}. 

To understand the structure of improvement counterterms in the $T\bar T$-deformed field theories, we considered renormalization of the canonical stress-energy tensor and the $U(1)$ current in the case of scalar and Dirac fields. We found that, to linear order in the $T\bar T$ coupling, neither stress-energy tensor nor the $U(1)$ current are renormalized in the massless case. However, this is not true if the undeformed theory is gapped. Since the very definition of the $T\bar T$ deformation in the presence of curvature is obscure, these observations partially unravel the way a $T\bar T$ theory couples to gravity and add an extra incentive to the program of generalizing the $T\bar T$ deformation to the theories living on a curved background.

It would be interesting to extend the analysis of this paper to other quantum field theories. Of particular interest is the case of interacting CFTs.
While we leave this problem for future research, we would like to sketch a possible route towards this direction.

As an example of the $T\bar T$-deformed interacting CFT, let us consider the Wess-Zumino-Witten (WZW) model on the group manifold $G$ in the presence of $T\bar T$ deformation. In conformal gauge for the two-dimensional metric, the (anti-)holomorphic components of the stress-energy tensor can be written in terms of the Kac-Moody currents, $T(z) =\frac{1}{\kappa} j^aj^a(z)$, $\tilde T(\bar z)=\frac{1}{\kappa}\tilde j^a\tilde j^a(\bar z)$. Here the constant $\kappa$
is determined by the level of the WZW model, the index $a$ is summed over $n$ values, where $n$ is the rank of the group $G$,
and $j^a(z)$, $\tilde j^a(\bar z)$ represent the (anti-)holomorphic Kac-Moody currents. 
These currents exhibit simple correlation functions $\langle j^a(z)j^b(0)\rangle = \delta^{ab}/z^2$,
$\langle \tilde j^a(\bar z)\tilde j^b(0)\rangle = \delta^{ab}/\bar z^2$, which can be thought of as correlators of the Noether currents $\partial\phi^a$,
$\bar\partial\phi^a$ associated with translation invariance in the space of $n$ free massless scalar fields $\phi^a$.

Now notice that the calculation in  section \ref{subsec:partial phi n}
for a single  free massless scalar is applicable to the case of $n$ decoupled scalars $\phi^a$.
In particular, the non-renormalization of the stress-energy tensor, in this case, rests on manipulating the integrals over the correlation functions
of the currents $\partial\phi^a$, $\bar\partial\phi^a$. Furthermore, since the Kac-Moody currents have the same correlation functions as $\partial\phi^a$ and $\bar\partial\phi^a$,  we can literally repeat the same arguments to conclude that the stress-energy tensor for an arbitrary WZW model is not renormalized in the presence of $T\bar T$ deformation.

Finally, we would like to point out a possible application of our results in the context of entanglement entropy calculations for a $T\bar T$-deformed field theory.  In general, the flow of entanglement entropy along the one-parameter family of theories defined by (\ref{DiffEI}) can be formulated in terms of the correlation function of the renormalized $T\bar T$ operator and a modular Hamiltonian associated with the entangling region of interest \cite{Rosenhaus:2014nha}. This correlation function can be  calculated perturbatively in $\lambda$, see \cite{Rosenhaus:2014woa,Rosenhaus:2014ula,Rosenhaus:2014zza,Faulkner:2014jva,Faulkner:2015csl}, provided that the energy cut off is sufficiently low. Our findings might be useful to explicitly carry out this sort of calculations.  In particular, a comparison with the proposed holographic duals can be done to understand better strongly coupled $T\bar T$ theories, see \textit{e.g.}, \cite{Chakraborty:2018kpr,Chakraborty:2020udr} and references therein for a related discussion.

\section*{Acknowledgements}  We thank John~Cardy and Vladimir~Rosenhaus for helpful discussions and correspondence. This work is partially supported by the Binational Science Foundation (grant No. 2016186), the Israeli Science Foundation Center of Excellence (grant No. 2289/18) and by the Quantum Universe I-CORE program of the Israel Planning and Budgeting Committee (grant No. 1937/12).

\appendix

\section{Master integrals}
\label{appendix:integrals}

In this appendix we calculate the divergent parts of the master integrals
\begin{align}
\label{master integral general definition}
\I_{\alpha_1\beta_1\cdots \alpha_k\beta_k}
=\int d^dx\,f(|x|)\,\frac{x_{\alpha_1} x_{\beta_1}\cdots x_{\alpha_k} x_{\beta_k}}{|x|^{2(d-1+k)}}\,,
\end{align}
where $k=0,1,2,\dots$, and $f(|x|)$ is a test function introduced to regularize spurious IR divergence
(which is an artifact of expanding our integrands around $x=0$).
Imposing the condition $f(0) = 1$ we ensure that presence of the test function $f(|x|)$ does not
affect the form of the UV divergencies of the master integrals, which we are interested in.
In this paper we define 
\begin{equation}
d = 2 - \epsilon
\end{equation}
and for the purpose of regularization we are interested in terms which are singular in the $\epsilon \rightarrow 0$ limit.

For instance, in the case $k=0$ in (\ref{master integral general definition}),
integrating by parts, expanding around $\epsilon = 0$, and keeping only divergent term, we find
\begin{align}
\label{I0 expression}
{\cal I} &= \int d^dx\,f(|x|)\,\frac{1}{|x|^{2d-2}} \\
&= \frac{2\pi^\frac{d}{2}}{\Gamma\left(\frac{d}{2}\right)}\,\frac{1}{\epsilon}\,
\int d|x|\,\left(\frac{d}{d|x|}\left(f(|x|)\,|x|^\epsilon\right)-
\frac{df}{d|x|}(1+{\cal O}(\epsilon))\right)\notag\\
&\rightarrow \frac{2\pi}{\epsilon}\,.\notag
\end{align}
Similarly, we calculate
\begin{align}
\label{I2 expression}
\I_{\alpha\beta} &= \frac{\pi}{\epsilon}\, \delta_{\alpha\beta}\,,\\
\label{I4 expression}
\I_{\alpha\beta\mu\nu} &=\frac{\pi}{4\epsilon}\,\left(\delta_{\alpha\beta}\delta_{\mu\nu}
+(\beta\leftrightarrow\mu)+(\beta\leftrightarrow\nu)\right)\,,\\
\label{I6 expression}
I_{\alpha\beta\mu\nu\rho\sigma}&=\frac{\pi}{24\epsilon}\,\left(
\delta_{\alpha\beta} (\delta_{\mu\nu}\delta_{\rho\sigma}
{+}(\nu\leftrightarrow\rho){+}(\nu\leftrightarrow\sigma))
{+}(\beta\leftrightarrow\mu)
{+}(\beta\leftrightarrow\nu)
{+}(\beta\leftrightarrow\rho)
{+}(\beta\leftrightarrow\sigma)\right)\,.
\end{align}

\section{Details of $[(\bar\psi\psi)^n]$ calculation}
\label{appendix: details of bar psi psi to n}

Expanding around $d=2$ and keeping only singular terms in (\ref{Delta 2 2 bar psi psi intermediate}) we obtain
(where $c_2$ and $c_3$ are real-valued, hence the coefficient in front of them is doubled,
while $c_1$ is defined to explicitly contain its value plus the complex conjugate)
\begin{align}
\label{Delta 2 2 bar psi psi n calculation}
\Delta_{(2)}^{(2)}(\bar\psi\psi)^n&=\frac{\lambda n(n-1)}{64\pi^2}\,(\bar\psi\psi)^{n-2}\,(c_1+2c_2+2c_3)\,,\\
c_1&=-2\partial_\alpha(\bar\psi\gamma^\lambda\gamma^\rho\psi(0)\bar\psi(0)
\gamma_\nu\gamma_\lambda\partial_\rho\psi)\,\I^{\alpha\nu}
+4\partial_\alpha(\bar\psi\gamma^\lambda\gamma_\mu\psi(0)
\bar\psi(0)\gamma_\nu\gamma_\lambda\partial_\rho\psi)\,\I^{\alpha\mu\nu\rho}\\
&-2\partial_\alpha(\bar\psi\gamma^\lambda\gamma^\rho\psi(0)
\bar\psi(0)\gamma_\nu\gamma_\rho\partial_\lambda\psi)\,\I^{\alpha\nu}
+4\partial_\alpha(\bar\psi\gamma^\lambda\gamma_\mu\psi(0)
\bar\psi(0)\gamma_\nu\gamma_\rho\partial_\lambda\psi)\,\I^{\alpha\mu\nu\rho}+\textrm{c.c.}\,,\notag\\
c_2&=\partial^\lambda\bar\psi\gamma^\rho\gamma_\mu\psi(0)
\bar\psi(0)\gamma_\nu\gamma_\lambda\partial_\rho\psi\,\I^{\mu\nu}
+
\partial^\rho\bar\psi\gamma^\lambda\gamma_\mu\psi(0)
\bar\psi(0)\gamma_\nu\gamma_\lambda\partial_\rho\psi\,\I^{\mu\nu}\,,\\
c_3&=\frac{1}{2}\partial_\alpha\partial_\beta(\bar\psi\gamma_\lambda\gamma_\rho\psi(0)
\bar\psi(0)\gamma^\lambda\gamma^\rho\psi)\,\I^{\alpha\beta}
-\partial_\alpha\partial_\beta(\bar\psi\gamma_\lambda\gamma_\rho\psi(0)
\bar\psi(0)\gamma_\nu\gamma^\rho\psi)\,\I^{\alpha\beta\lambda\nu}\\
&-\partial_\alpha\partial_\beta(\bar\psi\gamma_\lambda\gamma_\nu\psi(0)\bar\psi(0)
\gamma^\lambda\gamma_\rho\psi)\,\I^{\alpha\beta\nu\rho}
+2\partial_\alpha\partial_\beta(\bar\psi\gamma_\lambda\gamma_\mu\psi(0)
\bar\psi(0)\gamma_\nu\gamma_\rho\psi)\,\I^{\alpha\beta\lambda\rho\mu\nu}\notag\\
&+\frac{1}{2}\partial_\alpha\partial_\beta(\bar\psi\gamma_\lambda\gamma_\rho\psi(0)
\bar\psi(0)\gamma^\rho\gamma^\lambda\psi)\,\I^{\alpha\beta}
-\partial_\alpha\partial_\beta(\bar\psi\gamma_\lambda\gamma_\rho\psi(0)
\bar\psi(0)\gamma_\nu\gamma^\lambda\psi)\,\I^{\alpha\beta\rho\nu}\notag\\
&-\partial_\alpha\partial_\beta(\bar\psi\gamma_\lambda\gamma_\nu\psi(0)\bar\psi(0)
\gamma_\rho\gamma^\lambda\psi)\,\I^{\alpha\beta\nu\rho}
+2\partial_\alpha\partial_\beta(\bar\psi\gamma_\lambda\gamma_\mu\psi(0)
\bar\psi(0)\gamma_\nu\gamma^\lambda\psi)\,\I^{\alpha\beta\mu\nu}\,.\notag
\end{align}
Using ${\cal O}(\lambda^0)$ e.o.m. one can show that
\begin{align}
c_1=4c_2=4c_3=\frac{32\pi}{\epsilon}\,\partial_\mu(\bar\psi\psi(0))
\partial^\mu(\bar\psi(0)\psi)\,.
\end{align}
Using this expression in (\ref{Delta 2 2 bar psi psi n calculation}) we obtain (\ref{Delta 2 2 bar psi psi n result}).

Finally, expanding around $d=2$ and keeping only singular terms in (\ref{Delta 2 3 bar psi psi intermediate}) we obtain
(this expression is real-valued, so adding complex conjugate to it simply doubles its value)
\begin{align}
\label{Delta 2 3 bar psi psi n calculation}
\Delta_{(2)}^{(3)}(\bar\psi\psi)^n
&=-\frac{\lambda n(n-1)}{64\pi^2}\,(\bar\psi\psi)^{n-2}\,\\
&\times\left(
\bar\psi(0)\gamma_\mu\gamma_\lambda\partial_\rho\psi(0)
\bar\psi(0)\gamma_\nu\gamma^\lambda\partial^\rho\psi(0)\,\I^{\mu\nu}
+\bar\psi(0)\gamma_\mu\gamma_\rho\partial_\lambda\psi(0)
\bar\psi(0)\gamma_\nu\gamma^\lambda\partial^\rho\psi(0)\,\I^{\mu\nu}
\right.\notag\\
&-\left.
\partial_\alpha(\bar\psi(0)\gamma_\mu\gamma_\lambda\partial_\rho\psi
\bar\psi(0)\gamma^\lambda\gamma^\rho\psi)\,\I^{\mu\alpha}
+2\partial_\alpha(\bar\psi(0)\gamma_\mu\gamma_\lambda\partial_\rho\psi
\bar\psi(0)\gamma_\nu\gamma^\rho\psi)\,\I^{\alpha\mu\nu\lambda}
\right.\notag\\
&-\left.
\partial_\alpha(\bar\psi(0)\gamma_\mu\gamma_\lambda\partial_\rho\psi
\bar\psi(0)\gamma^\rho\gamma^\lambda\psi)\,\I^{\mu\alpha}
+2\partial_\alpha(\bar\psi(0)\gamma_\mu\gamma_\lambda\partial_\rho\psi
\bar\psi(0)\gamma_\nu\gamma^\lambda\psi)\,\I^{\alpha\mu\nu\rho}
\right.\notag\\
&+\left.
\frac{1}{2}\partial_\alpha\partial_\beta(\bar\psi\gamma_\lambda\gamma_\rho\psi(0)
\bar\psi\gamma^\lambda\gamma^\rho \psi(0))\,\I^{\alpha\beta}
-\partial_\alpha\partial_\beta(\bar\psi \gamma_\lambda\gamma_\rho\psi(0)
\bar\psi\gamma^\lambda\gamma_\nu \psi(0))\,\I^{\alpha\beta\rho\nu}
\right.\notag\\
&-\left.
\partial_\alpha\partial_\beta(\bar\psi\gamma_\lambda\gamma_\mu\psi(0)
\bar\psi\gamma^\lambda\gamma^\rho \psi(0))\,\I^{\alpha\beta\mu}_{\quad\;\;\;\rho}
+2\partial_\alpha\partial_\beta(\bar\psi \gamma_\lambda\gamma_\mu\psi(0)
\bar\psi\gamma^\lambda\gamma_\nu \psi(0))\,\I^{\alpha\beta\mu\nu}
\right.\notag\\
&+\left.
\frac{1}{2}\partial_\alpha\partial_\beta(\bar\psi\gamma_\rho\gamma_\lambda\psi(0)
\bar\psi\gamma^\lambda\gamma^\rho \psi(0))\,\I^{\alpha\beta}
-\partial_\alpha\partial_\beta(\bar\psi \gamma_\rho\gamma_\lambda\psi(0)
\bar\psi\gamma^\lambda\gamma_\nu \psi(0))\,\I^{\alpha\beta\rho\nu}
\right.\notag\\
&-\left.
\partial_\alpha\partial_\beta(\bar\psi\gamma_\rho\gamma_\mu\psi(0)
\bar\psi\gamma^\lambda\gamma^\rho \psi(0))\,\I^{\alpha\beta\mu}_{\quad\;\;\;\lambda}
+2\partial_\alpha\partial_\beta(\bar\psi \gamma_\rho\gamma_\mu\psi(0)
\bar\psi\gamma^\lambda\gamma_\nu \psi(0))\,\I^{\alpha\beta\mu\nu\rho}_{\quad\quad\;\;\;\lambda}
\right.\notag\\
&-\left.
\partial_\alpha(\bar\psi\gamma_\lambda\gamma_\rho\psi(0)\partial^\lambda\bar\psi
\gamma^\rho\gamma_\nu\psi(0))\,\I^{\alpha\nu}
+
2\partial_\alpha(\bar\psi\gamma_\lambda\gamma_\mu\psi(0)
\partial^\lambda\bar\psi\gamma^\rho\gamma_\nu\psi(0))\,\I^{\alpha\mu\nu}_{\quad\;\;\;\rho}
\right.\notag\\
&-\left.
\partial_\alpha(\bar\psi\gamma_\lambda\gamma_\rho\psi(0)\partial^\rho\bar\psi
\gamma^\lambda\gamma_\nu\psi(0))\,\I^{\alpha\nu}
+
2\partial_\alpha(\bar\psi\gamma_\lambda\gamma_\mu\psi(0)
\partial^\rho\bar\psi\gamma^\lambda\gamma_\nu\psi(0))\,\I^{\alpha\mu\nu}_{\quad\;\;\;\rho}
\right)+\textrm{c.c.}\notag
\end{align}
Simplifying this expression using  ${\cal O}(\lambda^0)$ e.o.m. one obtains (\ref{Delta 2 3 bar psi psi n result}).

\end{document}